\documentclass[12pt,preprint]{aastex}

\shorttitle{Silicon carbide single-crystal absorption spectra}
\shortauthors{Hofmeister et al.}

\begin{document}
%
%
\title{Optical constants of silicon carbide for astrophysical applications.
II. Extending optical functions from IR to UV using single-crystal absorption spectra}

\author{A. M. Hofmeister, K. M. Pitman\altaffilmark{1}}
\affil{Department of Earth and Planetary Sciences, 
           Washington University, St. Louis, MO 63130, USA}
           \email{hofmeist@levee.wustl.edu} 
    
\author{A. F. Goncharov}
\affil{Geophysical Laboratory, Carnegie Institution of Washington, 
	Washington, DC 20015, USA}         

     \and

\author{A. K. Speck}
\affil{Department of Physics and Astronomy, University of
           Missouri-Columbia, Columbia, MO 65211, USA}
             
\altaffiltext{1}{currently at Planetary Science Institute, Tucson, AZ, 85719}

\begin{abstract}
Laboratory measurements of unpolarized and polarized absorption spectra of 
various samples and crystal stuctures of silicon carbide (SiC) are presented 
from 1200--35,000\,cm$^{-1}$ ($\lambda \sim$\,8--0.28\,$\mu$m)
and used to improve the accuracy of optical functions ($n$ and $k$) 
from the infrared (IR) to the ultraviolet (UV).
Comparison with previous $\lambda \sim$\,6--20\,$\mu$m thin-film spectra 
constrains the thickness of the films and verifies that recent IR 
reflectivity data provide correct values for $k$ in the IR region.  
We extract $n$ and $k$ 
needed for radiative transfer models using a new ``difference method'', which
utilizes transmission spectra measured from two SiC single-crystals with 
different thicknesses. This method is ideal for near-IR to visible regions 
where absorbance and reflectance are low and can be applied to any material. 
Comparing our results with previous UV measurements of SiC,
we
distinguish between chemical and structural effects at high frequency.
We find that for all spectral regions, 3C ($\beta$-SiC) and the 
$\vec{E}\bot \vec{c}$ polarization of 6H (a type of $\alpha$-SiC) have almost 
identical optical functions that can be substituted for each other in modeling 
astronomical environments. Optical functions for $\vec{E} \| \vec{c}$ of 6H 
SiC have peaks shifted to lower frequency, permitting   
identification of this structure below $\lambda \sim4\mu$m. 
The onset of strong UV absorption for pure SiC occurs near 0.2\,$\mu$m, 
but the presence of impurities redshifts the rise to 0.33\,$\mu$m. 
Optical functions are similarly impacted.
Such large differences in spectral characteristics due to structural and 
chemical effects should be observable and provide
a means to distinguish chemical variation of SiC dust in space. 
\end{abstract}


%

\section{Introduction}

Silicon carbide (SiC) is of great interest in the context of dust in a  
variety of astrophysical environments and 
is the most well-studied type of presolar grain
\citep[][and references therein]{cn04,bernatowicz2}. 
Isotope measurements exist for thousands of presolar SiC grains (mostly of the 
cubic $\beta$ structural form) and suggest that they came 
from AGB stars and 
supernovae; the 
majority of meteoritic SiC grains 
are thought to have originated in circumstellar shells of carbon stars 
(C-stars).
Astronomical SiC was discovered  
via observations of a broad $\sim11\mu$m emission feature 
seen
in many 
C-star spectra 
\citep{hackwell,treffers} 
that
is associated with stretching of 
the Si-C bond.  
In the absence of strong infrared (IR) features from amorphous or 
graphitic carbon 
grains,
the presence and variations of the 11\,$\mu$m SiC feature provide a useful 
diagnostic for inferring the physical conditions 
in the circumstellar dust shells of C-stars
\citep[e.g.,][]{bar87,wil88,chkw90,goeb95,speck97,sloan98,speck05}.
Consequently, study of SiC may allow us to test theories of grain formation, 
evolution
and destruction in many astronomical environments.
\citet{thompson} showed that relating observational data on the SiC 
peak position, strength, and shape
to
the underlying continuum temperature 
did not yield useful information about circumstellar shell structure and 
evolution, and concluded that radiative transfer modeling is necessary to 
disentangle the dust shell parameters.
The ability to perform and achieve the latter has been limited by a 
lack 
of grain-size-independent optical 
functions (real and imaginary indices of refraction $n(\lambda)$ and 
$k(\lambda)$)
for SiC at all relevant wavelengths.

Studying SiC presents several opportunities for understanding cosmic dust. 
In particular,
the prevalence of SiC grains around carbon stars and in presolar samples 
suggests that some SiC grains survive the interstellar medium (ISM) 
and are incorporated into new planetary systems 
\citep[e.g.,][and references therein]{bernatowicz2}.
However, the abundance of SiC in the ISM determined in most studies is 
extremely low
\citep[e.g.,][]{whittet,kemper}, 
with only one study estimating abundances consistent with 10\% of Si atoms 
being bound in SiC \citep{min}. 
Because UV extinction is the main constraint in most ISM dust grain models, 
this disparity is connected to the high photon flux in the UV \citep{kmh,zubko04,wd01}, 
coupled with the fact that optical strengths for SiC in the UV 
vary among published studies \citep[cf.][]{phillip58,choyke57}.

In addition, recent Spitzer Space Telescope observations have shown that 
carbon stars are much more common in lower metallicity Local Group galaxies 
than in our own Galaxy \citep{lagadec07,lagadec08,zijlstra06,sloan08}. 
Detection of an SiC feature near 11 $\mu$m provides a diagnostic for the nature of these 
objects.
Furthermore, SiC is commonly found in extreme carbon stars and planetary 
nebulae (PNe) in the Large Magellanic Cloud 
\citep{stanghellini,bernardsalas,gruendl}, 
and not commonly found in their
Milky Way counterparts \citep{barlow83,rinehart02,casassus01, speck08}.
Consequently SiC presents a tool for understanding the effect of metallicity on
dust formation, evolution, and survival. Thorough investigation of these 
environments requires reliable optical functions for SiC with a broad 
wavelength coverage.

\citet[hereafter designated Part I]{pitman08} provided new mid- and 
far-IR reflectance spectra and grain-size-independent optical functions 
for several SiC samples and discussed the limitations of existing mid- to 
far-IR 
datasets on $n$ and $k$ for SiC (e.g., use of the 
commercially available hexagonal $\alpha$ rather than the space-relevant cubic 
$\beta$ structural forms of SiC, dependence on grain-size distribution of 
original laboratory sample measured, and the relationship between structure 
and spectral parameters from symmetry analysis).
In particular, Part I confirmed that previous SiC reflectance studies 
\citep[e.g.,][]{spitzer59,bh83} correctly assessed peak strengths 
and widths but found that peak positions were about 4 cm$^{-1}$ too low, due 
to the low resolution and lack of internal calibration in the older 
instruments used. 
%
%
Quantitatively analyzed IR reflectivity data  
(Part I) and previous laboratory studies of SiC, particularly at high 
frequency, provide much of the information required
for the applications discussed above, but 
improvements are needed in the following areas: 

\begin{enumerate}
\item{Published visible-UV spectroscopic data for $\beta$-SiC cover only part 
of the needed spectral range, and past visible-UV reflectance and absorbance 
datasets for $\alpha$-SiC are not entirely consistent with one another. 
For both  $\alpha$- and $\beta$-SiC, $n$ was obtained only for the visible 
\citep{thibault,phillip58,phillip60,choyke57}.
Furthermore, those data are not in electronic form and have been 
incorporated into previous derivations of optical functions via 
digitization, which provides limited spectral resolution and accuracy.}

\item{For SiC, the near-IR to visible region has low reflectivity ($R <$ 20\%) 
and low absorption coefficients ($A <$ 0.1~$\mu$m$^{-1}$).  
%
As a consequence, not only do primary reflections occur at the front surface, but 
moreover, secondary
reflections at the back surface of the sample are significant (Fig.~\ref{scheme}).
These back-reflections affect both reflection and absorption/transmission measurements, 
thereby contributing errors to the extracted optical functions for these regions.}

\item{In the UV region, both $R$ and $A$ are high, from which accurate 
optical functions can be extracted; however, few 
modern laboratories are equipped for 
measurements above about 33,000 cm$^{-1}$.
Very high frequency data in early literature can be evaluated and utilized if 
data are corrected for 
reflectivity are gathered in the visible.}
\end{enumerate}

To address these issues, this paper provides new, single-crystal 
absorption spectra of the same samples used in Part I, but at higher 
frequencies and over the spectral range where $R$ is low, to better constrain 
$k$ up to the UV. 
Combining these new data with the optical functions from Part I and our 
previous thin-film IR data \citep{speck99,speck05} both augments existing data 
by significantly extending the wavelength coverage, and provides a cross-check 
on the accuracy of the previous data.

In addition we develop a new approach -- the ``difference method'' -- 
for extracting $n$ and $k$ values from transmission spectra and also utilize
``limiting values'' as in previous literature \citep[e.g.][]{lipson60}. 
Our approach can be applied to any material, not just SiC.
Because transmission is easily and commonly measured, 
the difference method and limiting values may be used to 
greatly expand the database of astronomically relevant 
near-IR to UV optical functions and aid implementation of radiative transfer 
models for many environments, not just C-stars.

\section{Experimental methods}
\label{expmeth}

We investigated seven different samples, listed in
Table~\ref{samplelist}. Although the prevalent form of astronomical SiC is the cubic 
(3C, or $\beta$-SiC) polymorph, 
large single-crystals of that type are not available.
Hence, we examined vapor deposited 3C as well as 
several crystals with one specific hexagonal, layered 
structure (6H). The label $\alpha$ is used to describe diverse
hexagonal variations of the SiC structure. 
Part I describes sample characteristics in detail, 
which are 
summarized in 
Table~\ref{samplelist}. The colors of the samples are used to denote which sample is being examined.
Because color is indicative of impurities, this attribute is 
significant, as will be discussed in \S~\ref{color}.

The 6H SiC samples are anisotropic, so it is necessary to gather spectral data 
parallel and perpendicular to the layers (the $\vec{a}$ and $\vec{c}$ crystal 
axes) 
to provide data for the extraordinary and ordinary rays, respectively.
Sample thicknesses were determined by using a calibrated micrometer and also 
by using a binocular microscope with a calibrated reticule.
If possible, absorbance spectra were obtained from commercially polished 
sections.  Thin samples were prepared by grinding and polishing sections of 
moissanite, a gemstone grade variety of 6H SiC.  Twin planes provided 
reference points to maintain correct orientations ($\vec{a}$-$\vec{a}$ or 
$\vec{a}$-$\vec{c}$ plates).
For the near-IR to visible studies of tiny platelets, thicknesses were 
obtained from interference fringes using:

\[ d = \frac{1}{2n\Delta\nu} \]

\noindent
where $d$ is thickness and $\Delta$$\nu$ is the wavenumber spacing of the 
interference fringes.  Fringes tend to occur where $d$ is similar to the 
wavelength of the incident light and when the faces are perfectly parallel.  
Values of $n$ from our reflectivity study (Part I) and interference fringe 
spacing in the near-IR were used.  Nearly identical values of $d$ for thin 
platelets were obtained using direct measurements of $n$ from refractometry 
\citep{schaffernaum,schaffer} and fringe spacing in the visible.

Room temperature 
(18--19$^{\circ}$C) 
IR absorption spectra were acquired at 
normal incidence using an evacuated Bomem DA 3.02 Fourier transform 
spectrometer\footnote{Bomem Inc., Quebec, Canada.} 
(FTIR)
at 1\,cm$^{-1}$ resolution 
below $\nu$~$=$~4000\,cm$^{-1}$ and at 4\,cm$^{-1}$ resolution at higher 
frequencies.  Resolutions of this level are considered reasonable for 
non-gaseous samples acquired at room temperature.  The accuracy of the 
instrument is $\sim$0.01\,cm$^{-1}$. Far-IR data ($\nu < 650 {\rm cm}^{-1}$) 
were collected using a SiC globar source, a liquid helium cooled bolometer, 
and a coated mylar beam-splitter.
Mid-IR data ($\nu =$~450--4000\,cm$^{-1}$) were collected using a SiC globar 
source, a liquid nitrogen cooled HgCdTe detector, and a KBr beam-splitter.
Near-IR data ($\nu =$~1800--7000\,cm$^{-1}$) were collected using a SiC globar,
a liquid nitrogen cooled InSb detector, and a CaF$_{2}$ beam-splitter.
For $\nu =$~4000--10,000\,cm$^{-1}$, we used a quartz lamp, a liquid nitrogen 
cooled InSb detector, and a quartz beam-splitter. 
To access the 
$\nu =$~9500--21,000\,cm$^{-1}$ range, we substituted a silicon avalanche 
detector.
Large samples were mounted on apertures at the focal point in the sample 
compartment, and a wire-grid polarizer was used to obtain polarization 
information from the moissanite sample.
Most of our 6H samples were $\vec{a}$-$\vec{a}$ plates, from which unpolarized 
data (equivalent to the orientation $\vec{E} \bot \vec{c}$, 
where $\vec{E}$ is the electric field vector) were gathered.
Small platelets from Alfa/Aesar\texttrademark\, were placed on apertures of 100\,$\mu$m 
diameter and data were collected using a Spectra-Tech IR microscope\footnote{%
Spectra-Tech Inc. (Thermo Electron Corp.), Stamford, CT, USA.}.

Additional, independent measurements were obtained from
the platelets (on apertures) 
at UV wavelengths
using the spectroscopic facilities at the Geophysical 
Laboratory, Carnegie Institution of Washington. 
This system consists of a 
custom-made,
all-reflective 
microscope with Cassegrain-type mirror objectives and two off-axis paraboloids 
as a transfer optics, 
coupled to a Nicolet 750 Magna FT-IR 
spectrometer for IR measurements and to a grating monochromator-spectrograph 
with a CCD detector, 
and
configured for the visible and UV regions
\citep{goncharov06}.

We used a CaF$_{2}$ beam-splitter and a halogen lamp for near-IR acquisitions, 
and a combination halogen/deuterium light source for 
the
visible-UV spectral range.
Spectral resolution was about 16 cm$^{-1}$ through the whole spectral range 
(2400--40,000 cm$^{-1}$).

\section{Data Analysis and the Difference Method}
\label{dataanalysis}

Because transmission through a sample is more easily and commonly measured 
than reflection, we present two methods for extracting $n$ and $k$ from 
transmission measurements: the ``difference'' and 
limiting methods. Which method is applicable depends not only the 
material property $A$ 
(or equivalently $k$) but also on the thickness. Because optical
functions are frequency dependent, for any given thickness, each method will 
give accurate results 
over only certain spectral segments.
Practical issues such as sample characteristics (Table~\ref{samplelist}) 
strongly influence thickness and therefore
which method is viable for the region of interest.  In addition, the experimenter can vary thickness to 
optimize the accuracy of the results.
%
The equations underlying both methods are definitions for the various optical properties. 
The following analysis assumes normal incidence of a 
collimated light beam on a sample 
and that interference fringes do not exist or can 
be removed, e.g., by smoothing. Instruments record the light that is 
transmitted through a sample, rather than the light that is actually absorbed
by it.
Software for both spectrometers provide absorbance using the convention from 
the spectroscopy, chemistry and mineralogy literature (SCM),

\begin{equation}
a_{SCM} = - \log_{10}(\frac{I_{meas}}{I_{0}}),
\label{abseq}
\end{equation}

\noindent
where $I_{meas}$ is the intensity exiting the sample and $I_{0}$ is the 
intensity entering the sample (which is obtained by collecting a reference 
spectrum). The effect of surface reflections is neglected in the SCM 
literature largely because $R$ is unknown.  Instead, empirical baseline 
corrections are used, which suffice to extract widths and relative intensities 
needed for many applications. However, baseline corrections do not provide the 
information on $R$ and $n$ needed for modeling radiative transfer.

\begin{figure}
 \centering
 \includegraphics[width=16cm]{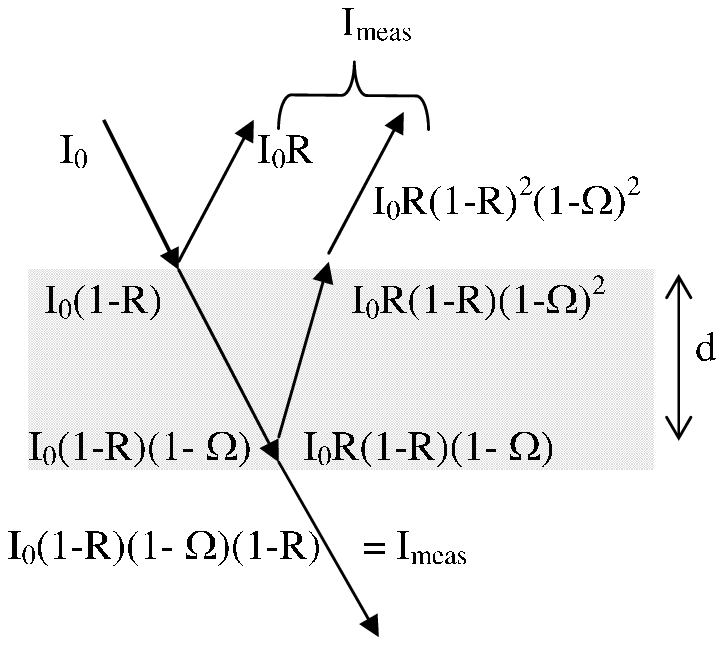}
\caption{Schematic of light loss in a dielectric material showing the most 
intense ray paths at near-normal incidence (the angle of incidence is 
exaggerated here to provide a clear representation of the processes).  
The incident light with intensity $I_{0}$ is partially reflected and 
transmitted at the front surface. The transmitted beam is partially 
absorbed, determined by thickness $d$, reflectance $R$, and absorptivity 
$= \Omega = I_{\rm abs}/I_{0}$.
At the back surface, the beam attenuated to $I_{0}$(1-$R$)(1-$\Omega$) is 
again partially reflected and transmitted.
For the case of a very weakly absorbing and thin sample, and if the angle 
of incidence allows the beam to reach the detector, then $I_{sum} = 2RI_{0}$ 
approximates the amount of light measured in a reflectance experiment.
In contrast, $R$ is typically 20\% for SiC in spectral regions other than 
the mid-IR, and should not be neglected in analyzing transmission data. After 
\citet{hofm03}.}
 \label{scheme}
 \end{figure}

To compute the amount of light that is actually absorbed, the true or ideal 
absorption coefficient ($A$) requires correcting $a_{SCM}$
for specular (surface) reflections of magnitude $R$, where $R$ is the 
proportion of incident light reflected only from the front surface as shown in 
Fig.~\ref{scheme}, as follows:

\begin{equation}
a_{SCM} = \frac{A(\nu)d}{2.3026} - 2\log_{10}(1 - R).
\label{lababseq}
\end{equation}

\noindent
Conversion to natural logarithms via Eq.~\ref{lababseq} is needed to directly 
compare measured absorption spectra with the ideal absorption coefficients $A$ 
obtained from reflectivity data (see, e.g., Part I, Figures 3a--10a).

Transmission spectra measured from two samples with different thicknesses, $d$,
but that are otherwise identical, are described by Eq.~\ref{lababseq} 
with different values for $d$.
Taking the difference between those equations (not shown) eliminates the 
reflection term, thereby allowing determination of $A$:

\begin{equation}
A = 2.3026\frac{a_{thck}-a_{thin}}{d_{thck}-d_{thin}}.
\label{Aeq}
\end{equation}

\noindent
where the subscripts ``thin'' and ``thck'' indicate the two different sections 
being measured. Specular reflectance $R$ is then obtained from:

\begin{equation}
R = 1-10^{(d_{thin}A/2.3026-a_{thin})/2}
\label{Req}
\end{equation}

\noindent
We refer to this approach as ``the difference method.'' To the best of our knowledge, 
this equation has not previously been published, although the above analysis is an extension
of existing analysis in the spectroscopy 
literature \citep[i.e., use of limiting values for transparent spectral regions, 
e.g.,][]{lipson60}, described below. 

The imaginary part of the complex index of refraction, $k$, is defined as
\begin{equation}
k = \frac{A}{4\pi\nu}.
\label{akeqn}
\end{equation}

\noindent
Since we can extract both $k$ and $R$ from the two transmission spectra, 
we can then obtain $n$, the real part of the complex index of refraction, 
from the following equation:

\begin{equation}
R \approx \frac{(n - 1)^{2}+k^{2}}{(n + 1)^{2}+k^{2}}
\label{rknueq}
\end{equation}

which assumes that the surrounding medium (air or vacuum for single-crystal 
studies) has $n$ = 1.

Applying this analysis requires meeting certain conditions. The thicker 
section cannot be opaque over any part of frequency range being investigated. 
The thinner section cannot be so thin that its absorbance is negligible. 
The polish must be nearly 
perfect for both sections and the thicknesses must be such that fraction of 
light reflected from the front surface is similar to the fraction of light 
transmitted through each crystal. Also, any experimental uncertainties 
connected with measuring thicknesses will be propagated into errors in the 
optical functions. Multiple measurements can decrease the uncertainties
associated with computing differences, provided that all sections are 
identical in composition and surface polish.

Due to the stringent requirements for the difference method, 
we utilize cross-checks.
One independent confirmation exists for $n$ 
in the visible because in this spectral region, $k \ll$1 for small to 
moderate values of $A$. For this case, Eq.~\ref{rknueq} simplifies to

\begin{equation}
R \approx \frac{(n - 1)^{2}}{(n + 1)^{2}}, 
\label{rnueq} 
\end{equation}

\noindent
which can be evaluated using measurements of $n$ in the visible, 
obtained independently by matching the sample's optics to that of a liquid 
with known refractive index.

A second cross-check, or alternative approach to our ``difference method'' is use of limiting 
values for determining $n$ and $k$.
For regions where $A$ is nearly zero, spectral measurement of a fairly thin 
sample provides $R$ because transmission, $T$ = $I_{meas}/I_{0}$, is given by 

\begin{equation}
T = \frac{1-(R - 1)^{2}e^{-Ad}}{1-R^{2}e^{-2Ad}},
\label{Teq} 
\end{equation}

\noindent
and can be simplified for the case of negligible absorption to

\begin{equation}
T \approx \frac{1-R}{1+R}.
\label{Tapproxeq}
\end{equation}

  \noindent
Substituting
Eq.~\ref{rnueq} into Eq.~\ref{Tapproxeq} yields $n$ \citep[e.g.][]{lipson60}.
Both reflections from the front and back surfaces of the sample section are 
accounted for (Fig.~\ref{scheme}). Use of limiting values requires a well-polished sample.
To obtain $A$ or $k$, additional 
transmission data is then acquired from a very thick sample, and the results 
for $R$ are used in Eq.~\ref{lababseq}.

If both sample thickness and $A$ are very large, a special case occurs. 
When scaling large $A$ obtained from 1 mm sections to thicknesses near 
1\,$\mu$m, the correction for reflectivity becomes negligible and $A$ can be 
directly computed. Again, a good polish is needed or scattering occurs which has a frequency dependence.

By applying both the ``difference'' and limiting methods, we can compute $n$ and 
$k$ from multiple transmission spectra of identical samples of different 
thicknesses. By comparing the results over frequency regions where both 
methods are applicable, 
we can construct a consistent set of optical functions
and ensure that surface polish is not affecting the values. 
Such a comparison requires laboratory measurements over a wide range of thicknesses. 
Furthermore, this approach can be applied to any material, not just 
SiC.

\section{Results and Comparison to Previous Studies}
\label{results}

\subsection{Near-IR absorption of 3C $\beta$-SiC}

\begin{figure}
 \centering
 \includegraphics[width=16cm]{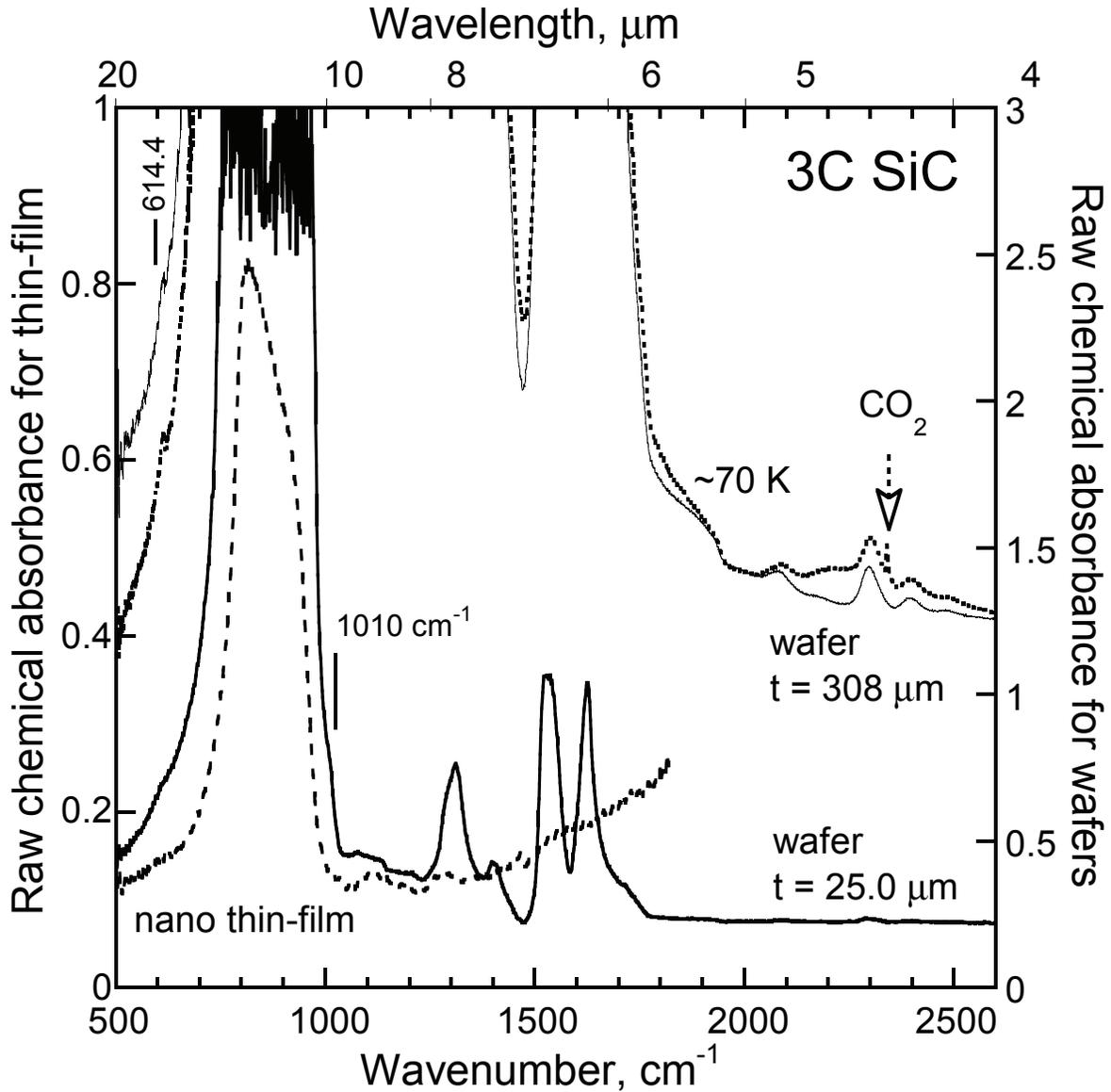}
  \caption{Raw absorption spectra of $\beta$-SiC.  Values are not 
corrected for any baseline or for reflectivity losses. 
 Thin 
solid line $\equiv$ 3C wafer with thickness $=$~308~$\mu$m.  
Thick dotted line $\equiv$ same sample at cryogenic temperatures.
Thick solid line $\equiv$ same wafer, thinned to 25~$\mu$m.  Dashed 
line $\equiv$ thin film of nanocrystalline $\beta$-SiC from 
Speck et al. (2005): only this sample pertains to the left side 
axis. Interference fringes exist in this spectrum.  Vertical lines 
denote weak peaks with positions given in wavenumbers.} 
 \label{betaabs}
 \end{figure}

Laboratory 
absorption spectra from a thin wafer of 
cubic SiC 
show one very strong IR mode from $\nu \sim$ 950--1000\,cm$^{-1}$ that is 
off-scale (Fig.~\ref{betaabs}). The noisy profile close to the 
barycenter is affected by stray light and is not intrinsic.  This main feature 
results
from Si-C stretching and is the transverse optic 
(TO) component. 
This triply degenerate mode also has a longitudinal optic (LO) component. 
Ideally, no other IR mode
is active in 3C.
However, as is common in diatomic substances \citep[e.g.,][]{hofm03}, 
structure is superimposed on this main band. A weak shoulder at
614.4\,cm$^{-1}$ is associated with the longitudinal acoustic (LA) 
mode, which is 
present 
due to resonance with the TO and LO modes.
Transverse acoustic (TA) modes have frequencies too low for such resonances to occur.
Peak positions of the fundamental modes are given in the footnotes in Table~\ref{table:1}.

From $\nu \sim$ 1000--1800\,cm$^{-1}$, a series of fairly intense
overtone-combination bands exist that are well resolved for sample thicknesses 
of $d \sim$~5--100~$\mu$m.  From $\nu \sim$ 1800--2700\,cm$^{-1}$, weak 
overtone-combination bands exist that are well resolved for 
$d \sim$~100~$\mu$m to several mm.
Positions of the overtones and the relationships to the fundamental modes are 
given in Table~\ref{table:1}.
Most of the modes are clearly connected to some combination of the 
fundamentals, but some cannot be assigned with confidence, 
as indicated by the question marks in this table. 
It is difficult to predict the activity and frequency of
overtone-combination modes because this type of phonon
involve anharmonic interactions of fundamental modes \citep[e.g.,][]{mitra}. 
Sometimes Raman modes can sum to provide overtones 
in the IR whereas some combinations of IR modes are forbidden. 
Overtones lacking an obvious assignment could be connected with impurities, 
non-stoichiometry, or stacking disorder.
 
By comparing intensities of the shoulder near 1800\,cm$^{-1}$ from the 
spectra of the thin wafer to spectra of a much thicker wafer, we confirmed 
that the thin wafer indeed has a thickness of 25~$\mu$m (Fig.~\ref{betaabs}).

Absorbance spectra at liquid N$_{2}$ temperatures of the 3C wafer 
were also acquired by attaching the thick 
wafer to a cold finger.  No shifts in peak positions
were seen. Bands are sharper but have roughly the same area as measured at 
room temperature (Fig.~\ref{betaabs}).  From this, we infer that absorption 
spectra of SiC are little affected by temperature, consistent with the 
incompressibility of SiC \citep[e.g.,][]{knittle}, i.e., bond lengths for 
silicon carbide vary little with pressure or temperature.
Positions and widths of the main bands will change even less than indicated by
Fig.~\ref{betaabs} upon cooling because $\nu$ of the fundamental mode 
changes at about half the rate of $\nu$ for the simple overtones.

The polish on the thin and thick sections were quite different (the vapor 
deposit has some texture) and opaque spectral regions exist, 
making it difficult to apply the difference method.
For example, near 2500\,cm$^{-1}$, using 
Eqs.~\ref{Aeq} and \ref{Req}
gives
$A$ as 0.00279~$\mu$m$^{-1}$ and $R$ as 0.048.
This computation of $R$ differs by a factor of 4 from the value in Part I 
and is 
attributed to
variations in polish on the surfaces of the samples.
We therefore determine the absorbance of the 1\,$\mu$m wafers 
(\S~\ref{composite3c})
by scaling from the very thick (308\,$\mu$m), strongly absorbing 
section.

\subsubsection{Composite spectra of 3C SiC; inference of thin-film thicknesses 
and band strengths}
\label{composite3c}

Thin film spectra of bulk $\beta$-SiC may not be precisely equivalent to 
spectra of the wafer because the bulk sample is not stoichiometric 
\citep{speck04} and because it was difficult to create uncracked 
thin films of uniform thickness for this extremely hard material.
 
\begin{figure}
 \centering
 \includegraphics[width=16cm]{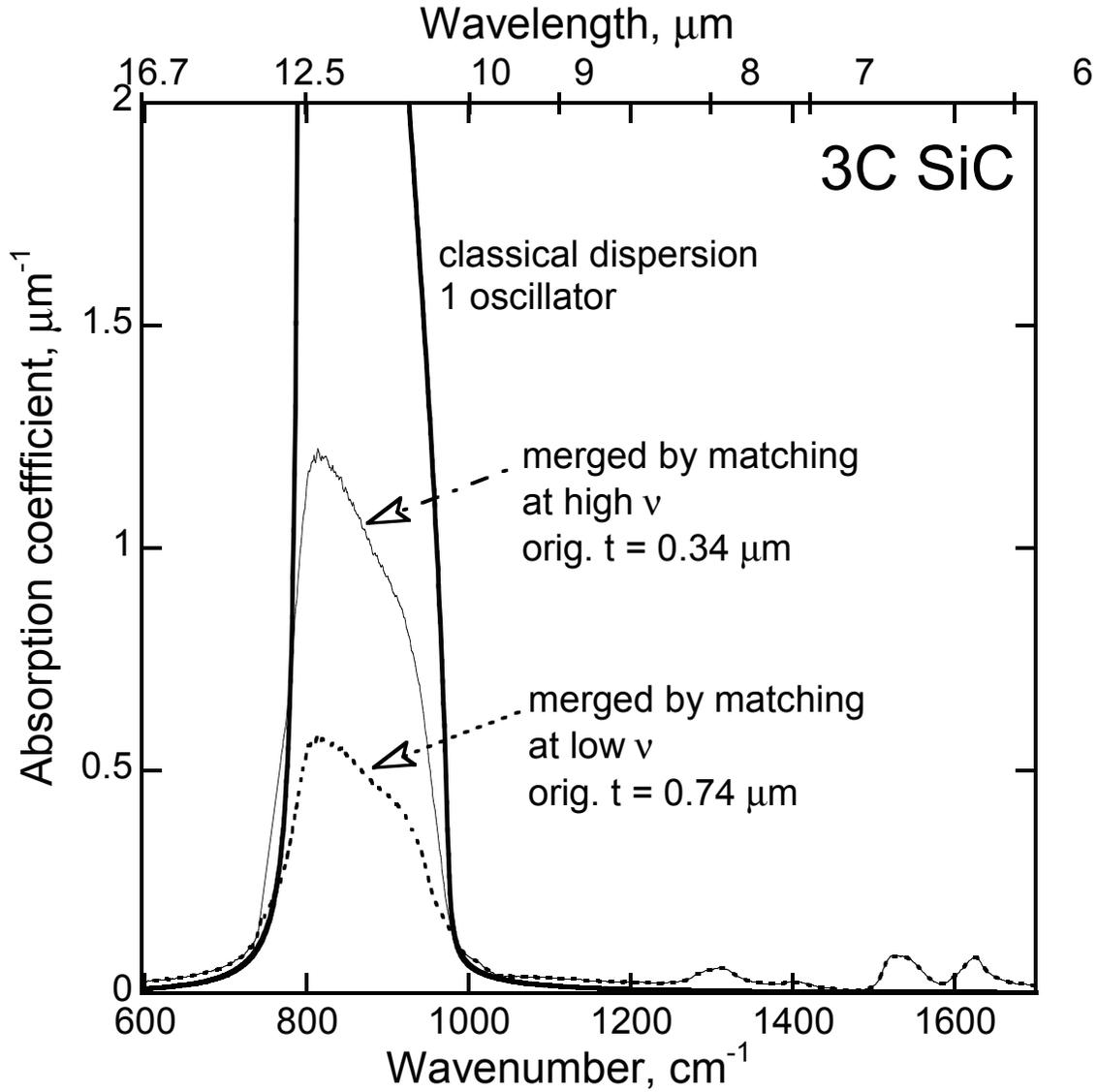}
    \caption{Baseline-corrected and merged absorption spectra 
of $\beta$-SiC.  Thin dotted line $\equiv$ merge to thin-film spectrum, 
scaled to match at high frequency.  The merge suggests an original film  
thickness of 0.74~$\mu$m.  Thin solid line $\equiv$ merge to thin-film
scaled to match at low frequency.  The merge suggests an original film 
thickness of 0.34~$\mu$m.  Thick solid line $\equiv$ ideal absorption 
coefficient calculated from classical dispersion analysis of measured 
reflectivity.} 
 \label{betacomp}
 \end{figure}

We instead compare absorption spectra of our wafer to thin-films of 
nanocrystalline $\beta$-SiC 
\citep[Fig.~\ref{betacomp}; data shown from][]{speck05}.  
In the thin-films, LO modes and shoulders are exaggerated due to a small 
proportion of the light crossing the film at non-normal incidence 
\citep[see][]{berreman}.
Because the reflections mainly arise from the smooth, highly polished diamond 
faces of the sample holder, 
rather than from the relatively rough thin-film, and are included in 
the reference spectrum, accounting for SiC reflections does not provide 
results that are consistent with reflectivity measurements and analyses 
\citep[see results for MgO by][]{hofm03}.
Therefore, we compared the raw thin-film data directly to the single-crystal 
measurements. We did not directly
compare our data to available absorption spectra from 
dispersions \citep[e.g.,][]{mutschke}, because peak profiles are affected to an
unknown degree by factors such as light leaking around grains 
\citep[e.g.,][]{hofm00a,hofm00b}.

The thin-film data could be matched to the ideal absorption coefficient at 
either high or low frequency, but not at both. Fig.~\ref{betacomp} therefore 
presents two possible merges for the SiC wafer and thin-film absorption 
spectra, one at 750\,cm$^{-1}$ and one at 1000\,cm$^{-1}$.
Although inferred thicknesses for both merges agree with results for other 
minerals \citep{hofm03,hb06}, the high frequency merge (suggesting an original 
thickness of 0.34~$\mu$m) is preferred because it gives more reasonable 
strengths for the overtones and is provided in electronic form 
(Table~\ref{betacomptab}). 
Overtones were not seen in thin-film spectra and must be sufficiently weak 
to be hidden by noise or interference fringes.
A thinner wafer is needed to reduce the ambiguity in merging spectra, thereby 
more accurately determining the thickness of the thin film, but it is 
difficult to grind samples to the ca. 5~$\mu$m in thickness needed while 
maintaining parallel faces.
Note also that our comparison of thin-film spectra to those of the wafer 
ignored the contribution of reflectivity as a function of frequency 
(see Eq.~\ref{lababseq}).

As a first approximation, the electronic file for the true absorption coefficient
 includes subtraction of 
0.00145 $\mu$m$^{-1}$ which was computed from the reflectivity correction term in Eq. 2 
using $R$ = 0.2 at 2500\,cm$^{-1}$. We neglect the frequency dependence of $R$ because 
for the thick, strongly absorbing samples the correction term is small.

\subsubsection{Laboratory measurements vs.\ ideal absorption coefficients}

Our merged $\beta$-SiC absorbance spectra can be reconciled with the ideal 
absorption spectra derived from classical dispersion analysis of 
reflectivity (Part I), as follows.  Due to light leaking through cracks in 
the thin-film, the main peak
and
$\nu \sim$ 800--950\,cm$^{-1}$
is rounded.
The merged absorbance spectra ``fits'' into the ideal absorption peak at about 
half-maximum.  Excess intensity near the LO position is expected, due to 
non-normal 
beam
incidence and possibly a wedge-shaped film (the diamond faces 
in the sample holder
are not perfectly parallel). The excess intensity from the LO component also 
shifts the absorption maximum from the TO component towards the LO position, 
as seen in the comparison to the high frequency merged spectrum 
(Fig.~\ref{betacomp}).
This comparison supports the lower value of the thickness (0.34~$\mu$m), 
inferred through comparison in the previous section.

\subsection{Near-IR absorptions of 
6H samples}

\begin{figure}
 \centering
 \includegraphics[width=10cm,angle=0]{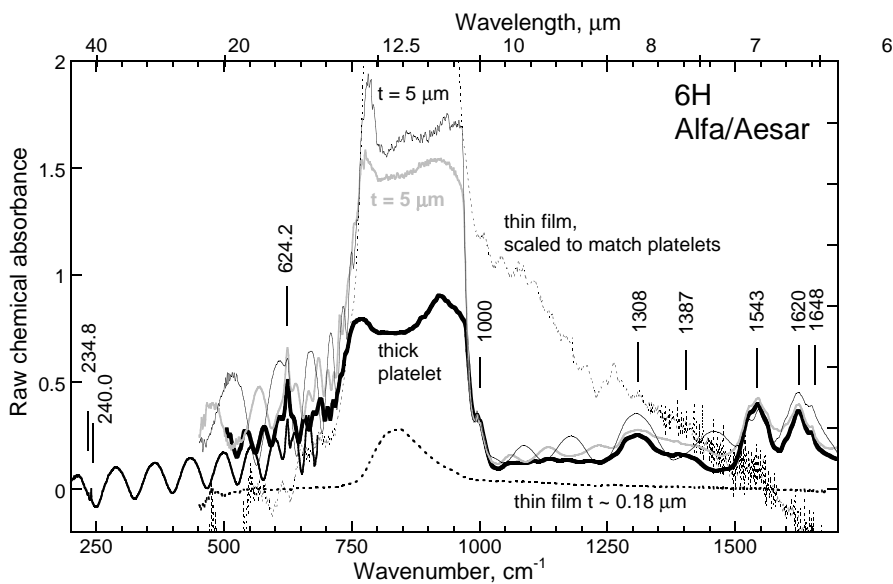}
   \caption{Absorption spectra of 6H SiC, $\vec{E} \bot \vec{c}$, 
from Alfa/Aesar. Except 
for the thin-film made from crushed platelets presented in 
\citet{speck05}, all spectra are from single-crystals. Thin line and gray 
line $\equiv$ very thin platelets with nominal thicknesses of 5~$\mu$m. 
Heavy line $\equiv$ platelet with a nominal 
thickness of 15~$\mu$m which was scaled to match spectra of the 5~$\mu$m
platelets. The scaling used suggests that the sample actually was originally 22~$\mu$m thick. 
Dotted line $\equiv$ thin film.  From scaling (light dotted line), 
the thickness of the film is 0.38~$\mu$m.  
Because interference fringes exist in most spectra, intrinsic
peak positions are indicated by vertical lines with labels in cm$^{-1}$.} 
 \label{6alphacomp}
 \end{figure}

Figure~\ref{6alphacomp} compares measured absorbance spectra from 6H platelets 
of various thicknesses to a thin-film made from the same material.
The thin-film may contain some particles oriented with $\vec{E} \| \vec{c}$, 
but this situation is unlikely because crushing would flatten the platelets 
rather than reorient them. A small (below 5\%) component of $\vec{E} \| \vec{c}$ would not be 
detected.

Two platelets with measured thicknesses of 5~$\mu$m gave slightly different 
absorbance spectra.  Most of the 
differences are observed for the 
low intensity peaks and result from superposition of interference fringes on 
the weak peaks. Because the positions (spacing) and intensities of these
  interference fringes vary among the spectra taken from the three platelets, 
the weak overtone-combination modes can be discerned (Fig.~\ref{6alphacomp}).  
In addition, the far-IR data
show some low frequency modes associated with zone folding 
\citep[see Part I;][]{nakashima}.

Variations seen in the height of the main peak originate in 
slight differences in thickness that are within the uncertainty of the 
measurements. The 
flat 
peak%
tops occur because 
of diffraction around the particles
and because samples of this thickness are 
opaque at peak center, due to the combination of high reflectance and 
absorbance.
Also, the peak height appears lower for the thick plate 
because of the combined effects of surface reflectance and stray light.

\begin{figure}
 \centering
 \includegraphics[width=16cm,angle=0]{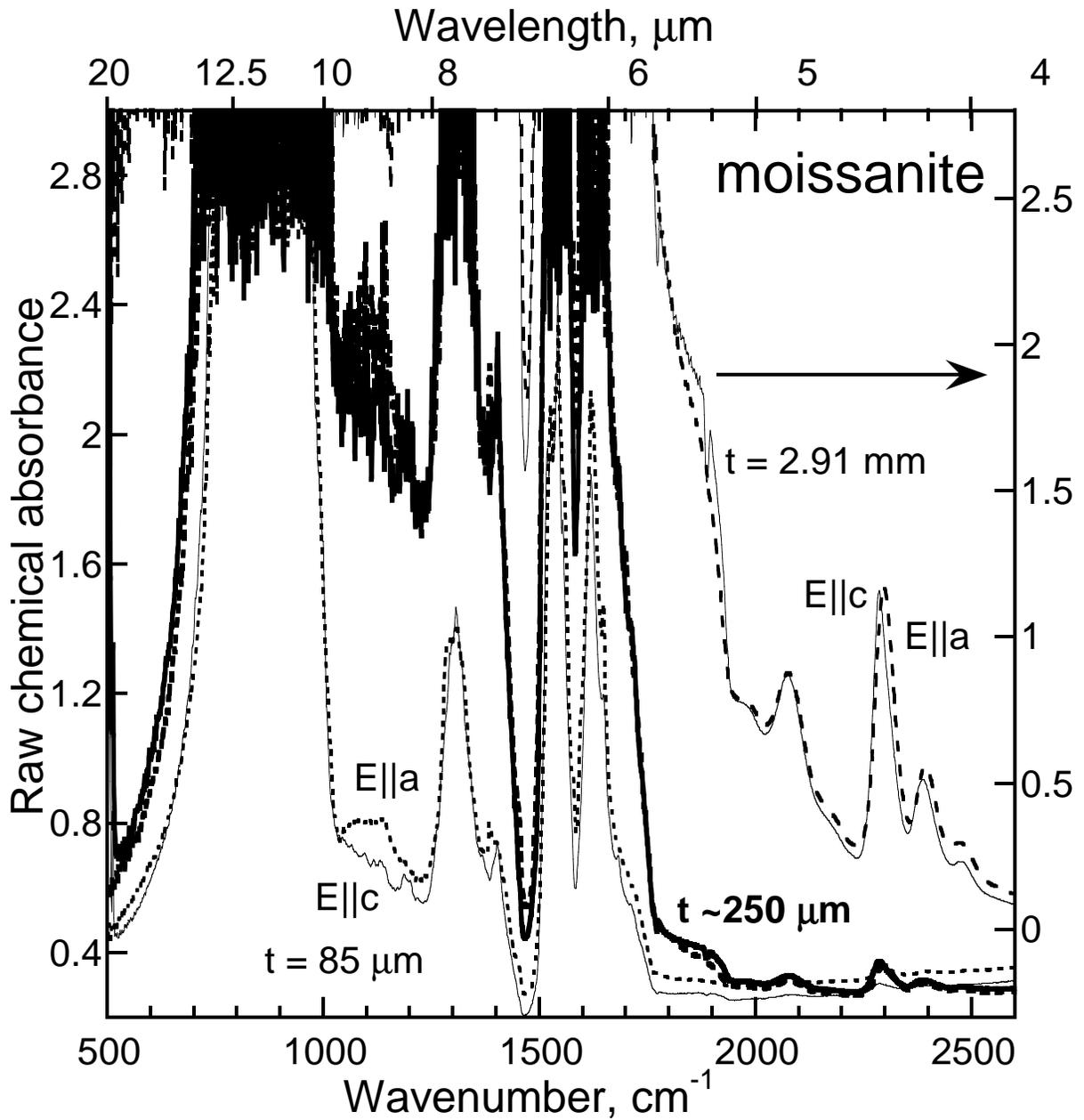}
    \caption{Polarized absorption spectra of moissanite (6H-SiC).  
Broken lines $\equiv$ $\vec{E} \bot \vec{c}$.  
Solid lines $\equiv$ $\vec{E} \| \vec{c}$. 
For the thinnest sample (light lines), the measured thickness was 85~$\mu$m.
The moderate thickness sample (heavy lines) was rougly 250~$\mu$m thick. 
Thickness is well-controlled for the largest sample 
(medium lines on the right-hand side, corresponding to the right-hand scale.)} 
 \label{mois3thick}
 \end{figure}

We also thinned 
the sample of moissanite and obtained absorption spectra for both 
polarizations for several thicknesses (Fig.~\ref{mois3thick}).  
For the SiC platelets and moissanite, peak positions appear to be the same 
within uncertainties of 1-2\,cm$^{-1}$, depending on peak width. We therefore 
report peak positions in Table~\ref{table:1} obtained from the best resolved 
peaks in the moissanite spectra. 
Assignments are based on proximity of the peak positions obtained from summing 
the frequencies of the fundamental modes and on the shifts of peak positions 
between the two polarizations.  

%
Based on binocular microscope and micrometer measurements, the thickness of 
the thinnest moissanite piece is constrained to be 85--90\,$\mu$m.
In addition, the thickness of the thick platelet was uncertain because this 
sample was too small to measure using a micrometer, and the determination 
using the binocular microscope was suspect.
By comparing the average of the two absorbance spectra obtained from two thin 
SiC platelets to the absorbance spectrum obtained from the thick SiC platelet 
(Fig.~\ref{6alphacomp}), we infer that the latter had a thickness 
of 22~$\mu$m.  Comparison with moissanite spectra gave a similar value.
Note that Fig.~\ref{6alphacomp} shows a scaled spectrum, not the absorbance 
for a 22~$\mu$m section.

\subsubsection{Cross-checks: comparing 6H and 3C SiC}
\label{xchecks}

Hexagonal forms of SiC frequently have stacking faults and twinning
that could produce extra, internal reflections and affect extracted values
of $R$ and $A$. 
We therefore compared measured absorbance spectra 
of 3C SiC (Fig.~\ref{betaabs}) to those of our various 6H samples 
(Fig.~\ref{6alphacomp} and Fig.~\ref{mois3thick}). We found that 
peak positions and spectral profiles differed only slightly among 
the various orientations and layering arrangements, and thus internal 
reflections
are not a concern for the thinner samples. 

Similarity of the 3C to 6H spectra supports equivalence of 
spectra from the platelets and moissanite. The main difference in spectra occurs between the 
two polarizations of 6H and 
3C
largely results from the 10\,cm$^{-1}$ difference between the
TO peak positions of Si-C 
fundamental
stretch in the two orientations. Intensities and widths seem to not be 
strongly affected 
by structural form.
\subsubsection{Composite spectra of 6H SiC; inferred 
thin film
thicknesses 
and band strengths}

\begin{figure}
 \centering
 \includegraphics[width=12cm]{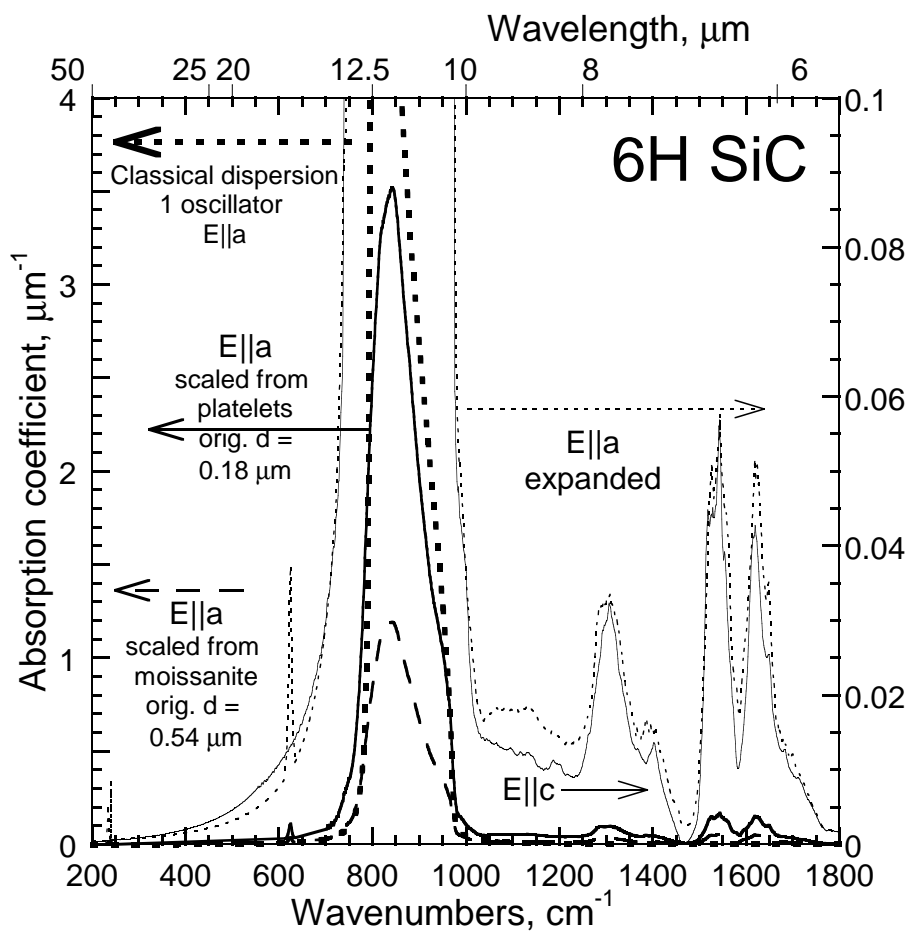}
    \caption{Merged polarized absorption spectra of moissanite at 
high frequency with the platelets at low frequency for $\vec{E} \bot \vec{c}$. 
Thick dotted line $\equiv$ ideal absorption coefficient calculated from 
classical dispersion analysis of measured reflectivity. 
Thick solid line $\equiv$ merged spectrum, scaled assuming 
platelet thickness of 5\,$\mu$m and film thickness of 0.18\,$\mu$m.
Dashed line $\equiv$ merged spectrum, scaled assuming largest moissanite slab 
was 2.91 mm thick, i.e., film was 0.54\,$\mu$m thick.
Fine solid line and right axis $\equiv$ expanded view of merge of moissanite 
at high frequency with classical dispersion analysis for $\vec{E} \| \vec{c}$. 
Fine dotted line and right axis $\equiv$ merged spectrum for 
$\vec{E} \bot \vec{c}$. Overtones are scaled based on the measured thickness 
of 2.91 mm.} 
 \label{composit}
 \end{figure}

Spectral segments 
with the highest resolution and least noisy peaks were merged to form the 
composite absorbance spectrum for $\vec{E} \bot \vec{c}$ over all wavelengths 
(Fig.~\ref{composit}), and for $\vec{E} \| \vec{c}$ at high frequency. 
The merge suggests an original thickness of 0.54\,$\mu$m for the thin film
and a thickness of 15~$\mu$m for the thinnest platelets.
The former value is consistent with the thickness inferred for the thin-film of the cubic SiC sample, 
whereas the latter value is outside the uncertainty of the determinations made with the binocular microscope.
The overestimation in the latter value largely results from the 
data containing specular reflections
that
effectively increase the 
absorption.  The fit within the peak envelope of the classical dispersion 
analysis is good.
If instead we assume that the thicknesses of the platelets are correct (i.e., 
we use an average of two thicknesses that were measured as near 5~$\mu$m), 
then the original film was 0.18~$\mu$m thick and fits within the envelope 
except near the shoulders. The different thicknesses obtained for the thin 
film through comparison is largely connected with reflectivity
not being subtracted.

The merged spectrum for the 0.54~$\mu$m-thick sample is provided in electronic 
form (Table~\ref{alphafinal}) for 6H because it is similar in thickness to the 3C film, 
and because we have only succeeded in making
films as thin as 0.18~$\mu$m for soft materials, which SiC is not.
For this spectrum, we spliced in segments in the far-IR from 
Paper I's classical dispersion analysis because our far-IR laboratory
absorption data had strong fringes. Fringes are unavoidable because the sample 
thickness needed to resolve the far-IR peaks is optimal for creating fringes 
in this range ($\lambda$=~25--100~$\mu$m). The fringes are very similar
to peak widths in this spectral region which makes spectral subtraction or 
polynomial fitting difficult to perform accurately.  Note that the
fits are very close on the edges of the peaks.
This spectrum can be multiplied by a factor of three, and still be within the 
uncertainties of our merge.  However, the overtone region is best represented 
by the data as provided.
For ideal values of $A$, the results from classical dispersion 
(electronic Table~\ref{alphaperp})
were substituted across the main peak.

For the $\vec{E} \| \vec{c}$ polarization (shown as light lines in Fig.~\ref{composit}), 
merged data are provided in
electronic Table~\ref{alphapara}.
For the main peak, synthetic spectra from classical dispersion analysis 
(Part I) were used to produce an ideal spectrum. 

\subsection{SiC absorptions in the visible to UV}

We studied
a relatively thick 6H platelet  
using both the Bomem (FTIR) 
spectrometer and Nicolet-CCD spectrometer system.  Data obtained from 
these two different instruments 
agree well
for frequencies below 16,000\,cm$^{-1}$ 
(not shown).  Above this frequency, high throughput and high sensitivity of available 
detectors overcome advantages of using an FT approach 
\citep[e.g.,][]{griffiths} and so we focus on results from the CCD 
spectrometer.

\begin{figure}
    \centering
 \includegraphics[width=12cm]{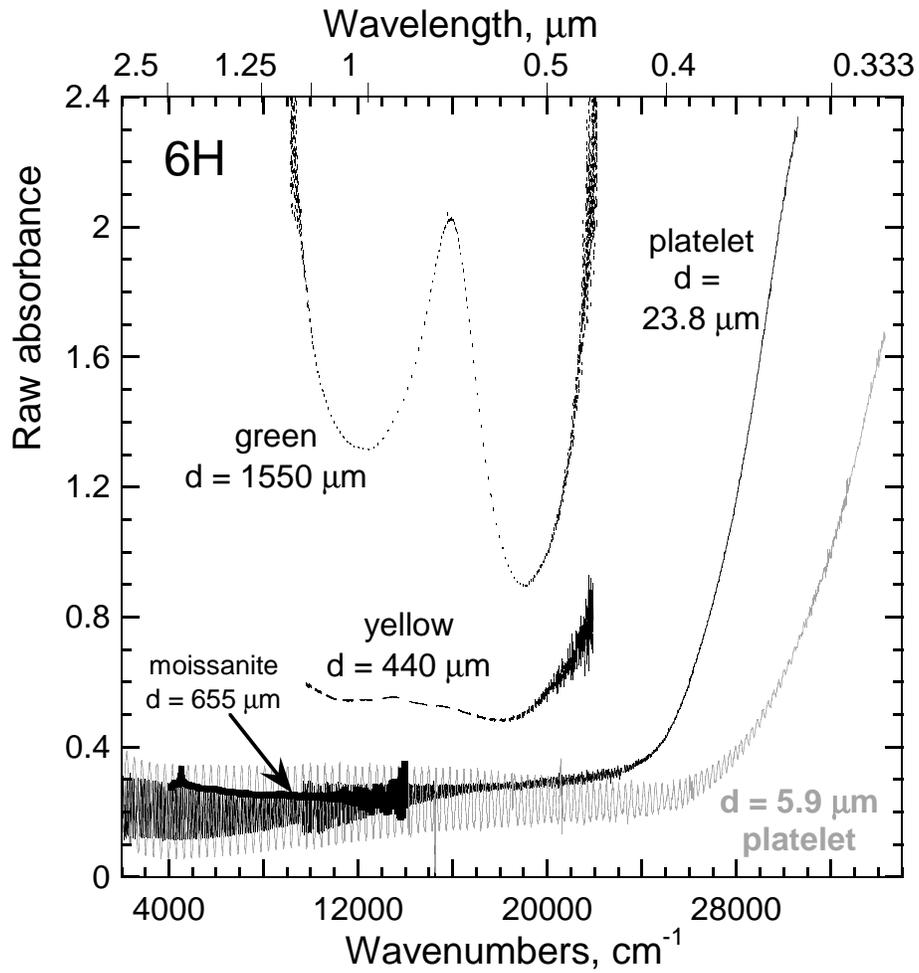}

    \caption{Unpolarized near-IR to UV spectra from 6H platelets and colored 
slabs.  
Fine black curve $\equiv$ thicker platelet.  
Gray $\equiv$ thinner platelet. 
Heavy black curve $\equiv$ small moissanite gemstone. 
Dashed line $\equiv$ yellow 6H slab.  
Dotted line $\equiv$ green 6H slab. 
Thicknesses as labeled.  
}
 \label{visplates}
 \end{figure}

Raw data obtained from two 6H
platelets with surfaces as grown
(Fig.~\ref{visplates})
contain interference fringes, the separation of which provides accurate 
thicknesses of 5.9 and 23.8~$\mu$m, respectively.
The absorbance of the thicker platelet is sufficiently high such that data 
acquisition was limited to $\nu < 30,500\,{\rm cm}^{-1}$.
To avoid fringes in our derived optical functions, we fitted the results to 
5th order polynomials above $\nu=20,000\,{\rm cm}^{-1}$ (not shown). 
To extrapolate data for the thick platelet up to $\nu =$35,000\,cm$^{-1}$, we 
linearly fit absorbance of the thick platelet from 30,000 to 30,600\,cm$^{-1}$. 
This linear extrapolation provides the lowest possible values of $A$ at the 
highest frequency attained, and the largest $R$ values (Fig.~\ref{visplates}).
The variation is small, however, and all such fits agree with $A$ obtained by 
\citet{phillip60} through Kramers-Kronig analysis of their UV reflectance 
data. Data on the platelets were also fit to higher (up to 9th) order 
polynomials and extended to lower frequencies, but the results 
(Fig.~\ref{visplates2})  
were
essentially the same.  In all cases, calculated 
reflectivity is consistent with direct measurements and index of refraction 
measurements at the red end of the visible, with a dip occuring near 
29,000\,cm$^{-1}$, and a steep rise to high frequency. There is some leeway in 
the fitting, which apparently underestimates $R$ and $n$ values and 
overestimates $A$ and $k$ in the calculations near 30,000\,cm$^{-1}$, and 
vice versa elsewhere.

\subsubsection{Stoichiometry, impurity, and color effects}
\label{color}

In this subsection, we use the observed spectral differences and the known 
connection of impurities with absorptions in the visible-UV region to 
ascertain trends.
Colors of samples generally indicated the presence of impurities. 
The colors of our samples are listed in Table~\ref{samplelist}.

The steep rise in $A$ and $R$ for the SiC platelets into the UV above 
$\nu =26,000\,{\rm cm}^{-1}$ is due to the band gap in this semiconducting 
material. 
The spectral characteristics appear to be affected by non-stoichiometry and 
other types of impurities existing in these materials, as indicated by 
coloration. 
For example, absorption measurements of 6H by \citet{phillip58} 
roughly agree with results by \citet{choyke57} for colorless samples, but their 
values are lower than extrapolation of our difference results for the tan 
platelets or with the trends for the yellow and green slabs. 

\begin{figure}
    \centering
 \includegraphics[width=8.0cm]{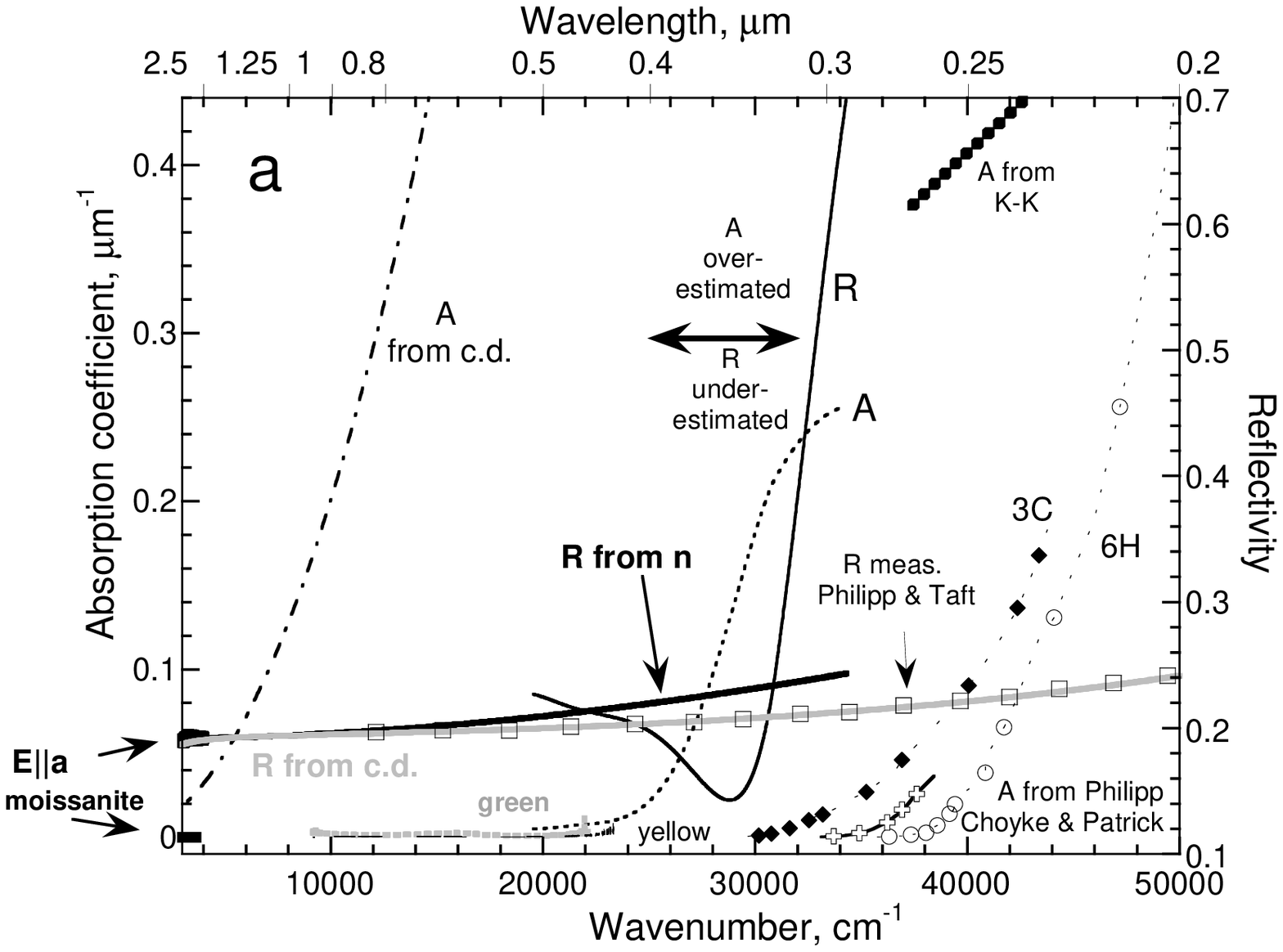}
\hfill
   \includegraphics[width=8.0cm]{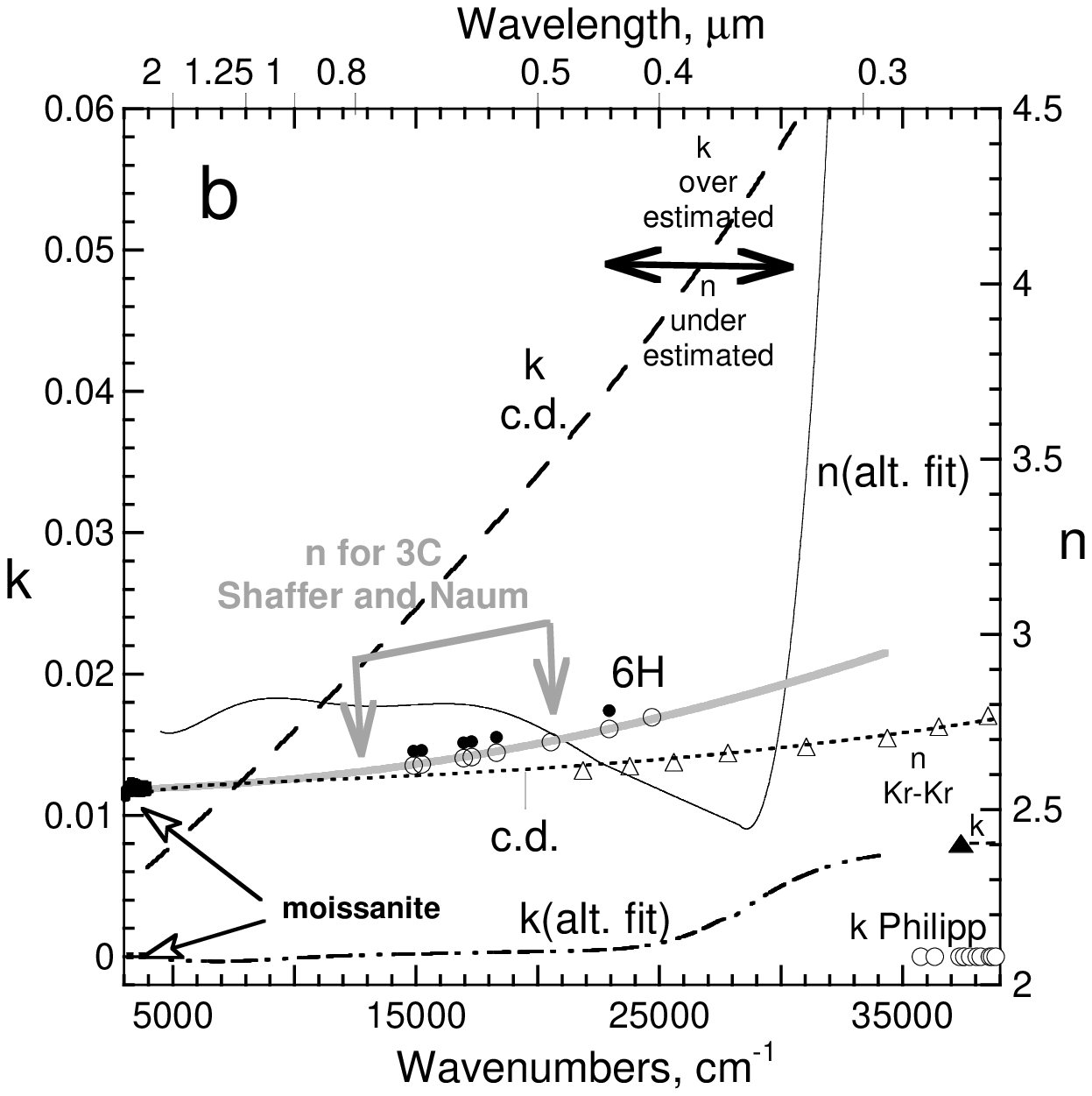}
    \caption{Properties derived from measurements in
Fig.~\ref{visplates} and comparison to previous data. (a) $A$ and $R$.
Dotted line $\equiv$ true absorption coefficient obtained by difference.  
Heavy double arrow denotes region where $R$ seems underestimated, see text.  
Light gray line $\equiv A$ for green 6H SiC calculated by using 
constant $R$ $\equiv$ 0.2. 
Dashed line $\equiv$ $A$ for yellow 6H calculated by constant $R$ of 
0.2. Heavy dotted line near 4000 cm$^{-1}$ $\equiv$ $A$ from near-IR 
measurements of the thickest moissanite slab. 
Diamonds $\equiv A$ measured for 3C by \citet{phillip58}. 
Circles $\equiv$  $A$ measured by \citet{phillip58}.
Open plus $\equiv A$ measured by \citet{choyke57}.
Dot-dashed $\equiv A$ from classical dispersion analysis of $R$ from 
by \citet{phillip60}.
Small squares $\equiv A$ from \citet{phillip60} using Kramers-Kronig analysis 
of $R$.
Solid line $\equiv$ reflectivity obtained by difference. 
Heavy solid line $\equiv$ reflectivity measured for moissanite (Part I). 
Heavy grey line = $R$ from classical dispersion analysis of reflectivity data 
(squares) from \citet{phillip60} which is indistinguishible from their 
Kramers-Kronig analysis. 
(b) Optical functions, mostly for 6H.  
Dot-dashed line $\equiv$ $k$ by difference using a high-order polynomical fit 
over all frequencies measured. 
Heavy dotted line $\equiv k$ from near-IR measurements of moissanite. 
Light solid line $\equiv n$ by difference using a high-order polynomical fit. 
Horizontal double arrow $\equiv$ region where polynomial fits were not perfect.
Heavy line $\equiv n$ from near-IR reflectivity data on moissanite of Part I.
Grey line $\equiv n$ from microscopy \citep{schaffernaum}, linearly 
extrapolated from the measured spectral region, denoted by arrows. 
Triangles filled $\equiv k$ and open $\equiv n$ from Kramers-Kronig analysis 
of reflectivity performed by \citet{phillip60}.
Dots $\equiv n$ from our classical dispersion fits to measurements of 
\citet{phillip60}.
Dashes $\equiv k$ from our classical dispersion analysis of \citet{phillip60} 
data.
Circles $\equiv k$ from absorbance measurements of colorless 6H by 
\citet{phillip58}.}
 \label{visplates2}
 \end{figure}

A steep rise to the UV for our deeply colored green and yellow 6H SiC begins 
near 21,000\,cm$^{-1}$ whereas spectra for the yellow 3C sample of 
\citet{phillip60} upturns near 32,000\,cm$^{-1}$, and spectra for their 
colorless samples upturn at 35,000--38,000\,cm$^{-1}$. The previous absorbance 
data were corrected for surface reflections using $n$ from microscopy. 
Although $R$ values from microscopy are 
%
slightly larger than the direct 
measurements (Fig.~\ref{visplates2}), 
small differences in $R$ give a negligible logarithmic correction term 
and are not important in determining $A$.
%
Also, the microscopy data were collected from 
colored SiC samples by \citet{thibault}, 
which have higher $A$ and $R$ than the 
colorless variety. The platelets are actually fairly dark, because color is 
apparent at 25 $\mu$m thicknesses, so it may not be valid to make a direct 
comparison to the previous data obtained from light colored samples 
(i.e. our darker samples contain more or different impurities). 
Given the steep rise for all samples,
 our linear extrapolation probably underestimated $A$ and 
overestimated $R$ above 31,000\,cm$^{-1}$ (Fig.~\ref{visplates2}), and the 
high-order polynomial fits are probably the better representation 
(Fig.~\ref{visplates2}). Nonetheless, the extrapolation in $A$ is consistent 
with that obtained from the Kramers-Kronig analysis of UV reflectance data for 6H SiC
by \citet{phillip60}.

Note also that $n$ of 3C  equals that of 6H SiC when light 
is polarized perpendicular to the $\vec{c}$ crystal axis; this is consistent with behavior at 
longer wavelengths. All evidence points to impurities, not 
structure, being the main cause of variation in the visible-UV region, as expected 
because electronic transitions are excited at such high frequencies.

\begin{figure}
 \centering
 \includegraphics[width=16cm]{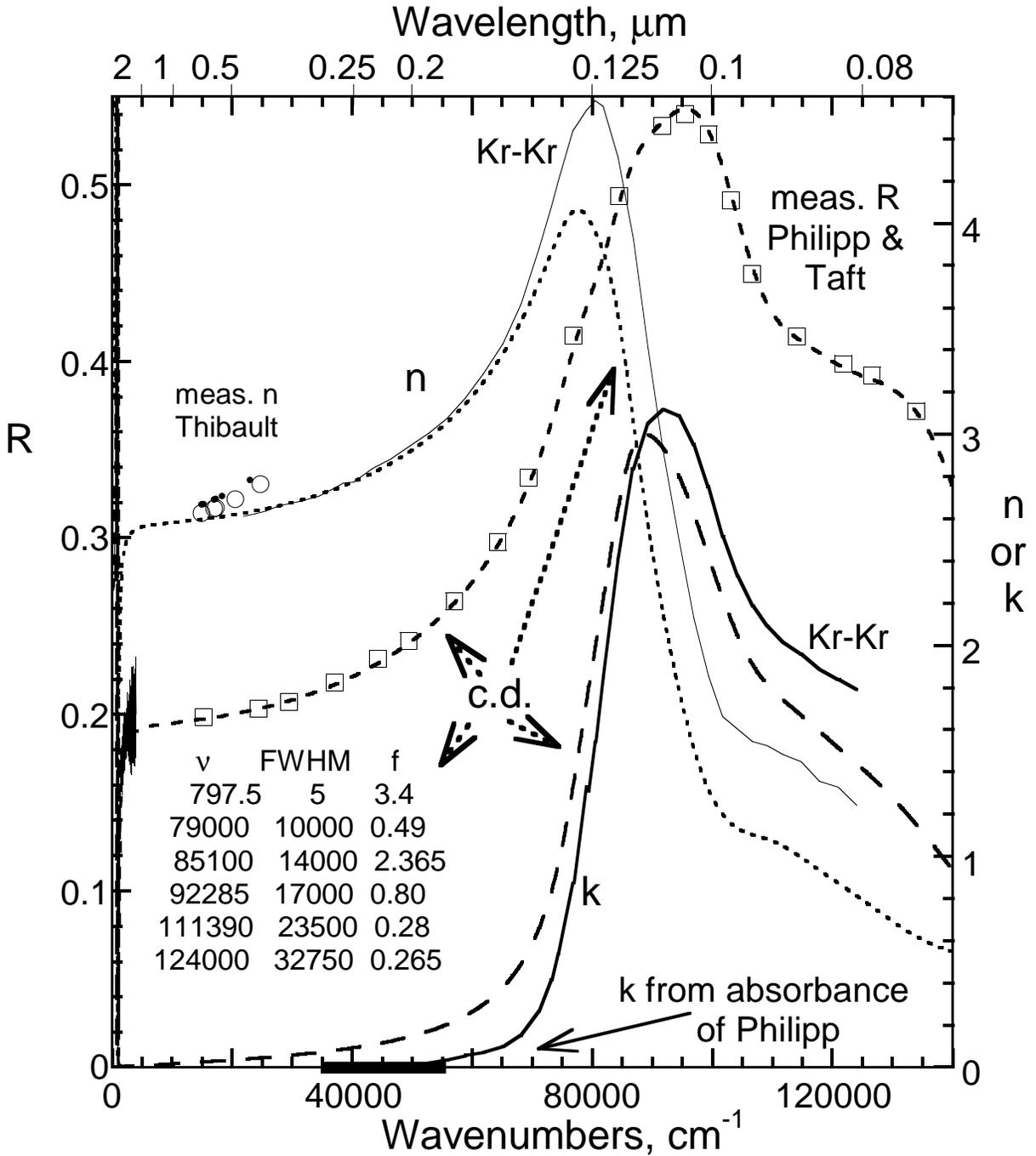}
    \caption{Classical dispersion analysis 
    (CDA)
    of UV reflectivity data from 
$\equiv$ $\vec{E} \bot \vec{c}$ of 6H SiC by \citet{phillip60}
Jagged line at low frequency is UV reflectivity of moissanite.
 Squares = measured values of $R$. 
Short dashes = CDA results for $R$.  
Dotted curve = CDA results for $n$.  
Long dashes = CDA results for $k$, 
based on the peak parameters listed in the inserted table. 
$\nu$ is peak position; FWHM is full width at half maximum,
both in cm$^{-1}$; $f$ is oscillator strength.
Solid lines = Kramers-Kronig analysis by \citet{phillip60} 
(thin for $n$, thick for $k$). 
Circles and dots are polarized measurements of $n$ for 6H SiC 
by \citet{thibault}
Thick black line = $k$ for the colorless 6H sample of \citet{phillip58}.
} 
 \label{classical}
 \end{figure}

The difference method (\S~\ref{dataanalysis}) calculates a decrease in $R$ between 
$\nu =$~23,000 and 29,000\,cm$^{-1}$, which is not in accord with classical dispersion analysis. 
This behavior is an artifact and is possibly due to 
the impurity content of the two platelets differing.
It is more likely that the surface smoothness differed slightly.
Variation in the amount of scattering 
is the most likely explanation for the dip in $R$. 
The steep rise in $R$, however, is attributed to impurities, given the 
correlation in the upturn in $A$ with the intensity of coloration. Reflectance 
will also upturn as $R$ and $A$ are correlated.  We propose that the impure 
samples have more band structure in the UV than does pure, stoichiometric SiC, 
and reflectance increases at lower frequencies for deep colors.  The linear 
extrapolation underestimates $A$ and overestimates $R$. This behavior and the 
trend in $n$ from microscopy of colored SiC suggests that impure, colored SiC 
samples have either an additional lower lying UV reflectivity band than the 
colorless variety, or perhaps that the entire band complex 
(Fig.~\ref{classical}) is shifted to lower frequency.

The similar raw absorbance obtained for both platelets and a much thicker 
moissanite sample (Fig.~\ref{visplates}) is consistent with absorption 
coefficient being essentially negligible in the near-IR to visible region. 
For this case, limiting values (see \S~\ref{dataanalysis}) are a 
better means of constraining optical functions than the difference method 
because errors result from subtraction of measured values of the absorbance 
that are very close to one another. First, absorbance and reflection 
measurements of the 2.91\,mm moissanite crystal 
constrain 
$A$ and $R$ near 4000\,cm$^{-1}$ (Fig.~\ref{visplates2}). Second, additional 
constraints are provided by the very thick green and yellow 6H slabs
(Fig.~\ref{visplates}).  The green 6H sample has a strong absorption feature 
near $\nu =$~16,000\,cm$^{-1}$ and the yellow 6H slab absorbs weakly in this 
spectral region.
Absorption coefficients were calculated for the green and yellow 6H slabs by 
using $R$ $=$ 0.2 
(Fig. 8, also see Part I), which is reasonable because $R$ varies 
little from $\nu =$~9000 to 23,000\,cm$^{-1}$.
Agreement is good with the difference calculation (Fig.~\ref{visplates2}), 
especially below 16,000\,cm$^{-1}$ where the Bomem system operates well.

\subsubsection{Classical dispersion analysis of previous UV reflectivity data}

Reflection measurements of 6H by \citet{phillip60} are consistent with 
a gradual increase in $R$ from moissanite measurements in the near IR.  
Kramers-Kronig analysis performed by \citet{phillip58} gives $A$ values that 
are consistent with our difference method but that are much larger than the 
measurements of \citet{phillip58}. Because of this discrepancy, we 
performed a classical dispersion analysis (see Part I) of \citet{phillip60}'s 
reflectivity data.
Using five oscillators with positions, FWHM and oscillator strengths ($f$) 
shown in Fig.~\ref{classical}, we were able to 
reproduce their measurements as well as our values for $R$ of moissanite in 
the near IR. We found that the best fits were obtained by setting 
$n_{\infty}$ = 1.56. Measurements of $n$ 
are made in the visible region and do not offer an independent constraint. 
Although good agreement was obtained with Kramers-Kronig analysis at low 
frequency for $n$, the results for $n$ are slightly lower than direct 
measurements of $n$ by microscopy (Fig.~\ref{visplates2}).
Values for $k$ do not agree. When overlapping peaks are present, as is the 
case here,  Kramers-Kronig analysis can fail, but the appearance of the 
optical functions does not suggest this is the case. Generally speaking, 
classical dispersion analysis is considered more trustworthy 
because it does not 
require extrapolation of $R$ from zero to infinity.
The classical dispersion analysis provides much higher values of $A$
(Fig.~\ref{visplates2}) than even the difference method. Such high values are 
not correct given the available absorption data, and point to small errors 
existing in the absolute measurements of $R$ 
at low frequency, i.e., where the detector was least 
sensitive.

As discussed in \S~\ref{xchecks}, stacking disorder occurs for hexagonal 
varieties of SiC. This presents a potential problem when measuring $R$ of 6H, 
because combined with small sample size it can lead to back reflections.
Whichever of these problems exists is manifest at low frequency, where $A$ is 
lowest.
Imperfections in surface polish are likely for the platelets.  This inflates
$R$ but not $A$ in the difference method. Errors in $R$ in the UV may also be 
instrumental 
effects
for these very difficult, vacuum measurements.
Absolute values for reflectivity also depend on the standard used for 
comparison (generally metallic silver).  

\subsubsection{Construction of high frequency optical functions for SiC}

This section combines the available data to provide two sets of high 
frequency optical functions for SiC, one that 
represents
stoichiometric, pure SiC, 
and another that represents impure SiC.  Both sets are valid for
the 3C structure 
and
6H with $\vec{E} \bot \vec{c}$.

The starting point at 4000\,cm$^{-1}$ for pure and impure SiC is provided by 
our measurements of $k$ and $n$ for the 2.91 mm thick sample of moissanite. 
For impure SiC, 
electronic Table~\ref{impure} 
is a linear
interpolatation of $k$ from the moissanite value at 4000\,cm$^{-1}$ 
to $k$ for the platelets at 20,000\,cm$^{-1}$ obtained from applying the 
difference method to the low order polynomial and incorporates values obtained 
from $A$ as shown in 
Fig.~\ref{visplates2}
to 30,000\,cm$^{-1}$.
The fit of \citet{schaffernaum}
constrains $n$ from 10,000 to 
30,000\,cm$^{-1}$ with linear interpolation below.
Above 50,000\,cm$^{-1}$, Table~\ref{impure} 
incorporates the classical dispersion 
analysis for $n$ and $k$. Between 30,000 and 50,000\,cm$^{-1}$, 
linear interpolations were made.

For pure SiC 
in electronic Table~\ref{pure},
$k$ between $\nu =$ 36,000 
to 56,000\,cm$^{-1}$ is
constrained using the absorbance of 6H $\alpha$-SiC 
measured by \citet{phillip58}, whereas classical dispersion analysis is used 
above 70,000\,cm$^{-1}$, and interpolation is made between the segments. For 
$n >$ 10,000\,cm$^{-1}$, classical dispersion analysis is used, and 
interpolation to moissanite data is made below 4000\,cm$^{-1}$.  Classical 
dispersion fits to \citet{phillip60} UV reflectance measurements of 6H provides for 
extrapolation to infinity.  For more accurate 
characterization of both impure and pure SiC, additional measurements of $n$ 
or of $R$ from gemmy, colorless moissanite are needed for both polarizations in the visible and UV.

\section{Discussion}

In carbon-rich dusty environments, SiC is one of only a few abundant dust 
species that exhibits an infrared spectral feature from which the 
physical parameters of these regions can be diagnosed 
\citep[see][and references therein]{speck08}. Consequently, SiC has
been of great interest to researchers seeking to understand the evolution of 
the dust shells and infrared features of carbon stars and their successor, 
C-rich pre-planetary and planetary nebulae. Furthermore, the lack of 
observable SiC spectral features in the ISM has lead to many studies which 
place upper limits on its abundance \citep[e.g.,][]{whittet,min,kemper}.
Moreover, the occurrence of presolar SiC grains provides the link between 
dust production in aging stars (AGB stars and supernovae) and subsequent 
formation of planetary systems, and thus SiC serves as a tool for dust studies 
in many environments.

Astronomical studies of C-rich environments utilize two forms of laboratory 
data: 
(1) absorbance spectra, which are often converted to mass absorption 
coefficients  
for direct comparison to observational spectra, under the assumption that the grain-sizes and shapes 
of the laboratory samples are similar to those dispersed in space, and 
(2) optical functions, which are use in various radiative transfer models
and in analyzing the effect of grain 
morphologies on spectral features, in which case single-reflection events are commonly assumed.
In \S~\ref{dataanalysis}--\ref{results},
we presented absorption coefficients for single-crystal SiC samples with both 
3C and 6H structural types and 
different amounts of impurities measured over a broader wavelength range. 
We focused on near-IR and visible frequencies because this region has been 
largely neglected. Merging our IR to UV measurements with previous far-UV data 
allowed us to provide continuous wavelength coverage for 
absorbance and optical functions of SiC, thereby permitting 
better constraints on its extinction 
for all astrophysical environments.
To provide $n$ and $k$ over all wavelengths, 
we expanded upon work by \citet{lipson60} and developed 
a method to extract optical functions from 
transmission measurements 
which to the best of our knowledge has not been used heretofore in astronomy. 
The basis is that reflection processes are an intrinsic part of 
transmission/absorption experiments (Fig.~\ref{scheme}).  
In Section 4, we compared our new data with 
previous measurements at high \citep[e.g.,][]{phillip60} and low (Part I) 
frequency, and found that values for 
optical functions are indeed high in the infrared. Through the comparison, we 
delineated
the strong effects of polarization and impurities. 
This section discusses limitations of the method 
to derive $n$ and $k$
and astronomical implications of the method and our results.

\subsection{Caveats in using difference method and limiting values to obtain $n$, $k$}

The difference method can be used to calculate $n$ and $k$ for any material from absorption 
measurements of crystals with different thicknesses.
However, accurate values are obtained only under certain conditions.  
Except for thickness, the samples used must be
identical: the orientation and impurity content must be the same, and the surface polish must be 
uniform for all samples. Imperfections in the surface polish do 
not seem to affect $A$ but alter $R$ from intrinsic values; this
affects the calculated $n$ strongly and $k$ to a minor extent.  
In addition, 
the measured absorbance cannot be minuscule. For thin samples where $A$ is near zero, limiting 
values provide better constraints.
Measuring absorbance of additional, larger crystals and obtaining independent measurements of the
$n$ will 
constrain $n$ and $k$ at all frequencies above the main peak.  
Using large, single-crystals is the most accurate means of determining $k$ for wavelength regions 
in which $k$ is low.  The difference method is particularly useful in the visible region, 
where absorbance is high but $k$ is low 
(see Eq.~\ref{akeqn}.) 

We found that the best approach is to collect data from single-crystals of 
diverse thicknesses and to use both limiting values and the difference method. 
For SiC, our analysis was complicated by the variation in color (impurities) 
for our samples and those previously studied. Color differences clearly alter 
absorption coefficients in the visible-UV region. We provide end-member values 
for $n$ and $k$, based on presence and absence of color.
The difference methods and limiting values cannot be applied to measurements 
of powder dispersions (see \S~\ref{powder}).

\subsection{Why $k(\lambda)$ derived from thin film, powder dispersion, 
reflectivity measurements differ}
\label{powder}
Although $k$ is a material property and should in principle be independent 
of the method used to determine it, in practice, values of $k$ for any given substance are 
influenced by the experimental technique.  The underlying problem is that laboratory studies do not 
directly measure absorption, but rather determine transmission which contains a reflection component,
and, moreover, at any given wavelength, $R$ and $A$ are correlated. 
Near the frequencies where light is strongly absorbed, 
light is also strongly reflected. For SiC in particular, $R$ 
approaches 97\% near $\lambda$ of 10\,$\mu$m.
Thus, at frequencies near the SiC fundamental mode, large particles are 
effectively opaque and the 3\% of light that is transmitted is attenuated 
sufficiently that the small amount which emerges from the sample
is within the noise level of detection.  For SiC, particles with sizes
near 1\,$\mu$m are effectively opaque. 
This size is consistent with grains in space and in powder dispersions used in 
laboratory studies. 
The effect creates non-linear dependence of transmission measurements, used to 
determine absorbance, on grain size for strong IR modes. 
For small-grain dipsersions (e.g., in a KBr pellet), 
the grains can clump, increasing the effective 
grain size to larger than the average created during grinding. 
This skews the grain-size distribution and makes a greater portion of the 
particles in the pellet 
effectively opaque,
which is problematic as optically thin conditions are
assumed in analyzing laboratory spectra.

Second, in addition to the processes of reflection and absorption, 
a non-negligible amount light is transmitted unattenuated through 
parts of the pellet matrix; 
powder dispersions transmit some unattenuated light at all frequencies 
because the particles do not completely cover the area. 
Such light leakage makes the fundamental peak in dispersion spectra of SiC 
both more rounded and lower in measured $A$ or derived $k$ values at peak 
center when compared to thin film spectra \citep[]{speck99,speck05} or in 
ideal absorption coefficients (Part I). 
For detailed discussion of light leakage \citep[see][]{hofm00b,hofm03}. 
SiC in particular is discussed by \citep[]{hofm00a}.  

The two effects combined mean that $A$ and thus $k$ (Eq.~\ref{akeqn})
obtained from powder dispersion methods is 
lower than the bulk grain-size independent property near the main peak 
obtained from reflectivity data. 
Measurements of $k$ from thin films are a little less than from 
reflectivity data, 
particularly at the main peak, but larger than $k$ from powder dispersions, 
because light leakage exists in the film, but is much less than in the 
dispersion.  In essence, it is the presence of reflectivity in all types of 
absorption measurements that causes the variation in the derived values of 
$k$ from intrinsic, reflectivity measurements.

These differences between measurement techniques preclude the use of powder 
dispersions with the difference method because
(1) the thickness of the material in a pellet sample is not well constrained; 
and 
(2) multiple reflections may occur from the particulates in the powder.

Consequently, thin film absorbance spectra offer an improvement over powder 
dispersion spectra \citep[e.g.,][]{hofm00b,hofm03}, but because film 
thickness is difficult to measure
accurately, quantities such as $k$, band strengths, and 
extinction coefficients 
are accompanied with experimental uncertainties.
In this paper, we constrained thickness of films previously 
used to collect mid-IR data through comparison with
near-IR spectra of single crystals.  The resulting high values of $k$ for
the main Si-C stretching mode support the results of Part I

The presence of reflections and light leakage in transmission measurements 
makes derivation of mass absorption coefficients from powder dispersion data
highly uncertain. This problem is exacerbated by the potential effect of grain 
shape.
%
Many astronomical studies use mass absorption coefficients obtained from 
powder dispersion data, $\kappa_{\rm abs}$, to estimate the relative 
abundances and total masses of dust species in a given circumstellar shell 
\citep[e.g.,][]{albrecht,stevens,kemper2,kemper}. In light of the limitations 
on using powder dispersions, the validity of these derived abundances is 
suspect.


Using the new optical functions presented here, it is possible to calculate 
 $\kappa_{\rm abs}$ for arbitrary grain sizes and grain shapes. Furthermore, 
the effect (or lack thereof) of impurities can be assessed. Thus the errors on
previous derivations of abundances and dust masses can be estimated and true
limits of these properties of dusty regions can be achieved.


\subsection{Importance of reflectivity in modeling extinction in the near-IR to visible regions}

In radiative transfer models, light that is extinguished by dust contains 
absorbing and scattering components. 
 
\[  Q_{\rm ext} = Q_{\rm abs} + Q_{\rm sca} \]

This section shows how the Q factors are related to material properties 
for the spectral regions focused on in the present paper and ties 
our results to radiative transfer modeling. 


The exact method by which $Q$-values are calculated varies, 
and the simplest case 
(spherical grains) uses Mie theory to compute these efficiency factors 
from the complex index of refraction $m = n - ik$ \citep{vandehulst}.
It is unlikely that astrophysical grains are all spherical, and recent studies 
have investigated the effects of grain shapes on spectral features 
\citep[e.g.,][]{bh83,mutschke,min2003,andersen2006}. 
However, the underlying relationship between the scattering and absorption 
efficiencies ($Q$ values) and laboratory measurements of $R$, $A$ from which 
$Q$s are derived will apply to all grain shapes. Here we use Mie theory as an 
illustration.

From Mie theory:
\begin{equation}
Q_{\rm abs} = - 4 x \Im \left( \frac{m^{2} - 1}{m^{2} + 2} \right)
\end{equation}

\begin{equation}
Q_{sca} = \frac{8}{3} x^{4} \Re \left( \frac{m^{2} - 1}{m^{2} + 2} \right)^{2}
\end{equation}
 
where $x = 2 \pi h / \lambda$ and $h$ is grain size. It is assumed that $x \ll 1$.

Using the definition $\lambda$ = 1/$\nu$ and Eq.~\ref{akeqn} we obtain:
\begin{equation}
Q_{\rm abs} = A\left( \frac{12nh}{(n^2-k^2+2)^{2} + 4n^2k^2} \right)
\end{equation}
From Eqn. 5 and Figures 8 and 9, $k$ is near zero
 for the transmitting near-IR to visible region. In this region $n$ is small but finite and not very frequency
dependent.  Consequently, the denominator in Eq. 12 is near the constant 4, and the numerator is 
proportional to the product $An$.  Through this relationship between 
$Q_{\rm abs}$ and $A$, 
the low values of absorptivity control values of $Q_{\rm abs}$ 
making it near zero in the near-IR to visible region.  Thus, extinction is due to scattering, 
not absorption, 
over this particular spectral region. 

To further probe the connection of radiative transfer with material properties, we simplify 
the expression for $Q_{\rm sca}$:
\begin{equation}
Q_{sca} = \frac{8}{3} (2\pi h \nu)^{4} \left( \frac{n^{2} - 1}{n^{2} + 2} \right)^{2}
\end{equation}
This is proportional to $R$ from eq.~\ref{rnueq}:
\begin{equation}
Q_{sca} = \frac{8}{3} (2\pi h \nu)^{4} R \left( \frac{ \left(n + 1 \right)^2}{n^{2} + 2} \right)^{2}
\end{equation}
Again, in the near-IR to visible spectral region, $n$ is small but finite and weakly dependent on 
frequency. To first order, the numerator and denominator involving $n$ cancel, and $Q_{sca}
$
in the near-IR to visible region is controlled
by its strong dependence on frequency and by the specific values of $R$. Our analysis shows that
reflection is important in radiative transfer, not only in spectral regions where 
absorption dominates, 
but even more so in spectral regions characterized by weak to negligible absorptions. 
Constraining $R$ in the visible through laboratory measurements is important.

\subsection{Potential uses of new data on $n$ and $k$ for SiC over the IR to UV}

In Part I we provided quantitative mid-IR laboratory measurements of SiC, 
which already improved the data available for astronomers in a variety of 
dusty fields, most notably by including $\beta-$SiC, rather than the 
$\alpha-$ polytype, and by having truly grain-size independent data.
However, the new data presented here augments Part I by extending the
wavelength coverage blueward. This is important because SiC grains are 
found in environments where the radiation field is strong in visible or UV 
regions.
Furthermore, our data show that while the main 11.3$\mu$m peak may not be 
diagnostic of grain impurities, the UV absorption is.

In addition, it is now possible to use the SiC optical functions to analyze 
grain-size and grain-shape effects
\citep[c.f.][]{bagnulo1,speck05}.
It has been suggested that there is an evolution in SiC grain sizes with the 
evolution of the star \citep{speck05}. 
Studies of the effect of grain size evolution have not been possible not only 
because of a lack of grain size independent data, but also because as the host 
stars evolve, the radiation field contains more high energy photons and the 
optical data needed for 
modeling has been lacking.
Furthermore, the SiC feature prevalent in carbon star 
spectra becomes much less common as these stars progress to the end of the AGB 
phase and into the post-AGB and PNe phases.

Whereas SiC is commonly seen in C-star spectra, its occurrence in
Galactic (Milky Way) planetary nebulae (PNe) is less common 
\citep[][]{barlow83,rinehart02,casassus01}, even though C-stars evolve into 
C-rich PNe. 
However, evolution of these grains 
from the C-star to PNe phase is not understood, and other spectral features 
(e.g., $\sim11-12\mu$m plateau due to PAH emission) may mask the SiC band 
\citep[see][]{speck04}.
Even though SiC is apparently rare in Galactic PNe - there are a handful of 
examples. We need to include SiC optical functions 
into the models to see how much can be hidden. 

This trend in the occurrence of the SiC feature from C-stars to pre-planetary 
nebulae and PNe is not echoed in recent studies of the LMC. The SiC feature 
remains observable, and thus including SiC into models of Magellanic cloud 
objects is essential.
Knowledge of optical properties of SiC is essential to understanding 
dust formation in lower metallicity environments such as the Magellanic Clouds 
(MCs) and local group galaxies 
\citep{lagadec07,lagadec08,zijlstra06,speck06,sloan08}.
Furthermore, recent observations of PNe in the Magellanic Clouds show that 
SiC features are much more common in these low metallicity environments 
\citep{stanghellini,bernardsalas}. 
In addition, the recent discovery of extreme (optically-obscured) C-stars in 
the LMC shows that SiC absorption features also occur more often in these low 
metallicity environments \citep{gruendl}.
Understanding why SiC features are more common in these highly evolved 
carbon-rich environments than their Galactic counterparts requires
reliable, grain-size-independent optical functions for SiC. 
Moreover, modeling of PNe requires wavelength coverage into the ultraviolet 
(UV) region, as this is where the radiation from the exciting central star 
peaks.

The fact that we can distinguish between samples of different purities based 
on the UV extinction has potentially far reaching consequences. 
In particular, the inclusion of SiC into dusty photo-ionization codes like 
MOCASSIN \citep{mocassin} and CLOUDY \citep{cloudy} is desirable. 
It will be very interesting to see the 
impact of varying SiC purity on the overall outputs of these codes.
Furthermore, the ability to use UV observations to determine the purity of the 
SiC grains provides a valuable test of the  hypotheses put forward by 
\citet{speck04} on the ``21$\mu$m'' feature, for example.

The changes in the UV absorption properties could potentially have huge 
impacts on hydrodynamic models of PNe shaping where the details of the 
interaction of light and dust grains is important.

\section{Conclusions}

We have discussed in detail the relationships between reflectance and 
transmittance measurements and how material properties 
such as optical functions can be extracted from 
such data. We emphasized that absorbance and absorption coefficients are 
derived, not measured, quantities, and that reflection plays a crucial role in 
both laboratory measurements and radiative transfer models.

Although absorbance can be compared directly to observational 
spectra in the IR, this does not take the effect of grain shape into account. 
The extracted $n$ and $k$ values can be used for more detailed 
analyses of astrophysical dust.

Reliable, grain-size/shape independent optical functions covering a broad 
wavelength range for a range of minerals is a growing need in the astronomy 
community.
To address these needs,

\begin{enumerate}
\item{
 we have devised a new 
``difference'' method, in order to obtain $n$ and $k$ from single-crystal 
absorption data.
For spectral regions where both reflectivity and absorptivity are low but 
non-negligible, transmission data from two sections with different, known 
thicknesses reflect the same amount of light at the front and back surfaces,
while simultaneously absorbing different amounts of light. 
The transmission data thus yield two equations with two unknowns, which can 
be solved uniquely for $n$ and $k$.
If used carefully, the difference method 
and use of limiting values 
can provide
$n$ and $k$ over a broad wavelength range for
astronomically relevant minerals using widely available transmission techniques.}

\item{We presented new SiC absorbance data which both extends the 
coverage of existing data from IR to UV wavelengths and 
checks optical functions derived from 
reflectance measurements. 
To improve upon our previous thin-film IR data \citep{speck99,speck05}, 
we merged it with new single-crystal spectra at slightly higher frequencies, 
thereby constraining film thickness \citep[after][]{hb06}
and allowing accurate extraction of $k$. 
This also makes direct comparison with 
absorption coefficients extracted from reflectivity data in \citet{pitman08}
possible.}

\item{We implemented  
the difference method to provide IR to UV optical 
constants for SiC ($\nu >$ 1000\,cm$^{-1}$) for 6H SiC with 
$\vec{E} \bot \vec{c}$.  
These data are also valid for 3C SiC, the $\beta$ structural type more 
dominant in astrophysical environments
but are affected by impurity content 
at high frequency. Infrared peak positions for $\vec{E} \| \vec{c}$ are 
shifted from those of 3C, particularly for the overtones, permitting  
the 6H species to be distinguished.
We have thereby extended our previous reflectivity 
determinations of $n$ and $k$ for SiC into the UV spectral region.
We also fitted available UV reflectivity to Lorentz oscillators as another 
check and to extend the frequency range.
}

\item{
We provided detailed measurements on SiC overtone-combination bands,
which are affected by structure and orientation.  We have shown that the 
impure (colored) and pure varieties of commercially prepared SiC produce identical $k$ 
values just below the visible.
We have also confirmed that the 3C spectrum is nearly identical to that of 6H 
when oriented with $\vec{E} \bot \vec{c}$ as determined in \citet{pitman08} 
for the fundamental frequency modes. 
Available data indicate that the equivalence holds for all spectral regions, 
if sample purity is the same.}

\end{enumerate}

Our new approach -- the difference method -- for extracting optical functions 
from transmission spectra can be applied to any material, not just SiC.
Because transmission is easily and commonly measured,  
our difference method may be used 
to greatly expand the database of astronomically relevant near-IR to UV 
optical functions and 
aid implementation of radiative transfer models for 
many environments, not just 
C-stars.


\acknowledgements
     Acknowledgements. This work was supported by NASA APRA04--000--0041 and 
     NSF--AST 0607418 and NSF--AST 067341.
     K.M.P. is supported by an appointment to the 
     NASA Postdoctoral Program, administered by Oak Ridge Associated 
     Universities.  
     A.F.G. acknowledges support from NSF/EAR, DOE/BES, DOE/NNSA (CDAC) and 
     the W. M. Keck Foundation
     The authors thank M. Meixner (STScI) for purchasing 
     the moissanite and the $\beta$-SiC wafer samples studied here, 
     and L. Valencic (NASA GSFC) for helpful conversations.

\clearpage

\clearpage

\clearpage

\clearpage

\clearpage

\clearpage

\clearpage   

\clearpage

\clearpage

\clearpage

\begin{table}
\footnotesize
\caption{Experimental samples descriptions}
\label{samplelist}
\renewcommand{\footnoterule}{}  
\begin{tabular}{c@{\hspace{2mm}}p{38mm}@{\hspace{2mm}}p{38mm}@{\hspace{3mm}}p{57mm}}     
\hline\hline
Polytype      & Mineral Name         & Grain Size       & Comments \\
 \hline\hline
$\alpha$-SiC & synthetic moissanite & diam. = 6.5\,mm    & round brilliant cut gem \\
$\alpha$-SiC & 6H-SiC  (amber) & 2\,$\mu$m powder; surf. area $=9-11~{\rm m}^{2}$\,g$^{-1}$ & hexagonal plates; purity: 99.8\% metals basis\\
$\alpha$-SiC & 6H-SiC  (blue)  & several mm per crystal& Intergrown crystals \\
$\alpha$-SiC & 6H-SiC (green) & diam. ca. 4~mm  & layered hexagonal crystals \\
$\alpha$-SiC & 6H-SiC (yellow) & diam. ca. 8~mm  & single crystal, some zoning\\
$\beta$-SiC & 3C-SiC\,wafer (fcc cubic) & diam$=$~5~$\mu$m &  CVD wafer purity: $\geq$~99.9995\% \\
$\beta$-SiC & 3C-SiC gray (fcc cubic) & diam. $\sim$~2.5--25~$\mu$m         & equant (spherical) chips\\
 \hline
 \hline
 $$
 \end{tabular}
 \end{table}

\clearpage


\begin{table}
\caption{Frequencies of overtone-combination modes in SiC, in cm$^{-1}$}
\label{table:1}      
\small
\centering                          
\begin{tabular}{c c c c c}        
\hline\hline                 
 & \multicolumn{2}{c}{3C}   & \multicolumn{2}{c}{6H} \\    
\raisebox{12pt}{Mode Type} & Calc. & Meas. & Calc.$^{\mathrm{a}}$ & Meas.$^{\mathrm{b}}$ \\ 
\hline                        
\hline
3TA~$=$~TO & 797 & 797.5 & 797 & 797.5 $\bot$ \\      
LA$+$TA & 881 & 881 & 891 & --- \\
TO$+$Folded & n.a. & 1012 & 1022, 1037$^{\mathrm{a}}$ & 1000 \\
TO$+$TA & 1064 & 1077b & 1054,1064 & 1073 $\bot$\\
LA$+$2TA & 1147 & 1125b & 1157 & 1136b \\
2LA & 1225 & 1207 & 1225 & 1194 \\
LO$+$TA & 1239 & --- & 1239 & --- \\
TO$+$2fold & n.a. & 1284 & 1272 $\bot$ & 1286 $\bot$ \\
TO$+$2TA & 1329 & 1311s & 1319, 1329 & 1308s \\
? & --- & --- & --- & 1387 $\bot$ \\
TO$+$LA & 1413 & 1402s & 1413, 1423 & 1404, 1412$^{\mathrm{a}}$ \\
2LA$+$TA & 1496 & --- & 1516 & --- \\
LO$+$TA & 1505 & 1525s & 1500, 1505 & 1520, 1526 \\
? & --- & 1535b & --- & 1543 \\
LO$+$LA & 1588 & --- & 1583,1588 & 1555,1560 \\
2TO & 1595 & 1625s & 1575, 1595 & 1618, 1620s \\
? & --- & --- & --- & 1648 \\
TO$+$TA$+$LA & 1678 & --- & 1678, 1688 & 1683, 1700 \\
2TA$+$2LA & 1762 & 1716 & 1782 & 1713,1720b \\
LO$+$TA$+$LA & 1854 & --- & 1860, 1864 & --- \\
2TO$+$TA & 1861 & 1920b & 1841, 1861 & 1876, 1890b,s \\
? & --- & --- & --- & 1897 $\|$s \\
TO$+$2LA & 2027 & --- & 2037, 2047 & 1980, 1990 \\
2TO$+$2TA & 2128 & 2079 & 2108, 2128 & 2076, 2080s \\
2TO$+$LA & 2209 & 2182b & 2210, 2219 & 2179b \\
TO$+$LO$+$2TA & 2302 & 2297s & 2287, 2297 & 2286, 2297s \\
3TO & 2392 & 2396 & 2364, 2392 & 2388, 2394s \\
2TO$+$LA$+$TA & 2475 & 2487 & 2465, 2485 & 2477, 2480 \\
\hline                                   
\hline
\end{tabular}
\begin{list}{}{}
\item[] Notes: b $\equiv$ broad, s $\equiv$ strong, n.a. $\equiv$ 
not applicable. 
``Calc'' values 
based on fundamental frequencies 
(Part I, 
Nakashima \& Harima 
2000), summarized as follows in $\nu$ (cm$^{-1}$) ($\lambda$ in $\mu$m = 
10,000/$\nu$):
TA = 266 in 3C, also at 234.8 and 244 in 6H; LA = 614.4 in 3C, 
or
624.4 in 
6H; TO = 797.7 in 3C, 
797.5 for $\vec{E} \bot \vec{c}$
, 
787.8 for 
$\vec{E} \| \vec{c}$; 
LO = 973 in 3C, 
975 for $\vec{E} \bot \vec{c}$ 
, 
970 for 
$\vec{E} \| \vec{c}$.
6H also has modes at 883.7 and 888.5 for $\vec{E} \| \vec{c}$.

\item[$^{\mathrm{a}}$]
For 6H calculated, if two entries are given, the lower frequency value 
pertains to $\vec{E} \| \vec{c}$.  
Modes marked $\|$ or $\bot$ are found only when $\vec{E}$ is oriented
parallel or perpendicular to the c-axis.
Unmarked entries indicate that 
the same frequency exists in both polarizations.
\item[$^{\mathrm{b}}$]
For the measured modes of 6H, modes in both polarizations are unmarked.  
If two entries are given, the lower 
frequency value pertains to $\vec{E} \| \vec{c}$; for these split modes, 
entries having lower values in $\vec{E} \| \vec{c}$ confirm involvement of 
the TO or LO components.
\end{list}
\end{table}
%


\clearpage


\begin{table}
\caption{Merged single-crystal absorbance spectrum of (3C) $\beta$-SiC$^\ast$}
\label{betacomptab}      
\small
\centering                          
\begin{tabular}{cc} 
\hline
Wavenumber & Absorption Coefficient \\
(cm$^{-1}$)& ($\mu$m$^{-1}$)\\
\hline                                   
401.12 &	-0.0054155\\				
402.09 &	-0.12127	\\			
403.05 &	-0.071841		\\		
404.02 &	-0.042450			\\	
404.98 &	0.70815				\\
405.94 &	0.035504				\\
406.91 &	-0.057883				\\
407.87 &	0.060957				\\
408.84 &	0.027782				\\
409.80 & -0.0026673	\\
\hline
\end{tabular}
\begin{tabular}{p{10cm}}
$^\ast$Table continued as electronic-only file.
\end{tabular}
\end{table}

\clearpage


\begin{table}
\caption{Merged absorbance spectrum of ($\alpha$) 6H-SiC, 
$\vec{E}\bot \vec{c}$ polarization (ordinary ray)$^\ast$}
\label{alphafinal}      
\small
\centering                          
\begin{tabular}{cc} 
\hline
Wavenumber & Absorption Coefficient \\
(cm$^{-1}$)& ($\mu$m$^{-1}$)\\
\hline                                   
3.0000	& 0.00020004\\
4.0000	& 0.00020007\\
5.0000	& 0.00020012\\
6.0000	& 0.00020017\\
7.0000	& 0.00020023\\
8.0000	& 0.00020030\\
9.0000	& 0.00020038\\
10.000	& 0.00020047\\
11.000	& 0.00020056\\
12.000	& 0.00020067\\
\hline
\end{tabular}
\begin{tabular}{p{10cm}}
$^\ast$Table continued as electronic-only file.
\end{tabular}
\end{table}

\clearpage


\begin{table}
\caption{Ideal absorbance spectrum of ($\alpha$) 6H-SiC, 
calculated from classical dispersion analyses, $\vec{E}\bot \vec{c}$ 
polarization (ordinary ray)$^\ast$}
\label{alphaperp}      
\small
\centering                          
\begin{tabular}{cc} 
\hline
Wavenumber & Absorption Coefficient \\
(cm$^{-1}$)& ($\mu$m$^{-1}$)\\
\hline                                   
3.0000 &	0.00020004 \\
4.0000 &	0.00020007\\
5.0000 &	0.00020012\\
6.0000 &	0.00020017\\
7.0000 &	0.00020023\\
8.0000 &	0.00020030\\
9.0000 &	0.00020038\\
10.000 &	0.00020047\\
11.000 &	0.00020056\\
12.000 &	0.00020067\\
\hline
\end{tabular}
\begin{tabular}{p{10cm}}
$^\ast$Table continued as electronic-only file.
\end{tabular}
\end{table}

\clearpage


\begin{table}
\caption{Ideal absorbance spectrum of ($\alpha$) 6H-SiC, $\vec{E}\| \vec{c}$ 
polarization (extraordinary ray)$^\ast$}
\label{alphapara}      
\small
\centering                          
\begin{tabular}{cc} 
\hline
Wavenumber & Absorption Coefficient \\
(cm$^{-1}$)& ($\mu$m$^{-1}$)\\
\hline                                   
2.0000 &	3.0000$\times 10^{-8}$\\
3.0000 &	7.0000$\times 10^{-8}$\\
4.0000 &	1.2000$\times 10^{-7}$\\
5.0000 &	1.8000$\times 10^{-7}$\\
6.0000 &	2.7000$\times 10^{-7}$\\
7.0000 &	3.6000$\times 10^{-7}$\\
8.0000 &	4.7000$\times 10^{-7}$\\
9.0000 &	6.0000$\times 10^{-7}$\\
10.000 &	7.4000$\times 10^{-7}$\\
11.000 &	9.0000$\times 10^{-7}$\\
\hline
\end{tabular}
\begin{tabular}{p{10cm}}
$^\ast$Table continued as electronic-only file.
\end{tabular}
\end{table}

\clearpage


\begin{table}
\caption{Complex refractive indices for impure SiC, 
applicable for either 3C ($\beta$) or  6H ($\alpha$) $\vec{E} \bot \vec{c}$ 
polarization (ordinary ray)$^\ast$}
\label{impure}      
\small
\centering                          
\begin{tabular}{ccc} 
\hline
Wavenumber &  $n$   & $k$ \\
(cm$^{-1}$)&        &     \\
\hline                                   
4000.0 &	2.5574 &	6.0687$\times 10^{-5}$\\
4001.0 &	2.5574 &	6.0676$\times 10^{-5}$\\
4051.0 &	2.5575 &	6.0984$\times 10^{-5}$\\
4101.0 &	2.5576 &	6.1292$\times 10^{-5}$\\
4151.0 &	2.5578 &	6.1599$\times 10^{-5}$\\
4201.0 &	2.5579 &	6.1907$\times 10^{-5}$\\
4251.0 &	2.5580 &	6.2215$\times 10^{-5}$\\
4301.0 &	2.5581 &	6.2523$\times 10^{-5}$\\
4351.0 &	2.5583 &	6.2830$\times 10^{-5}$\\
4401.0 &	2.5584 &	6.3138$\times 10^{-5}$\\
\hline
\end{tabular}
\begin{tabular}{p{10cm}}
$^\ast$Table continued as electronic-only file.
\end{tabular}
\end{table}

\clearpage


\begin{table}
\caption{Complex refractive indices for pure SiC, 
applicable for either 3C ($\beta$) or  6H ($\alpha$) $\vec{E} \bot \vec{c}$ 
polarization (ordinary ray)$^\ast$}
\label{pure}      
\small
\centering                          
\begin{tabular}{ccc} 
\hline
Wavenumber &  $n$   & $k$ \\
(cm$^{-1}$)&        &     \\
\hline                                   
4001.0 &	2.5501 &	6.0687$\times 10^{-5}$\\		
4051.0 &	2.5508 &	6.0591$\times 10^{-5}$\\		
4101.0 &	2.5515 &	6.0495$\times 10^{-5}$\\		
4151.0 &	2.5522 &	6.0400$\times 10^{-5}$\\		
4201.0 &	2.5529 &	6.0304$\times 10^{-5}$\\		
4251.0 &	2.5535 &	6.0209$\times 10^{-5}$\\		
4301.0 &	2.5542 &	6.0113$\times 10^{-5}$\\		
4351.0 &	2.5548 &	6.0017$\times 10^{-5}$\\		
4401.0 &	2.5553 &	5.9922$\times 10^{-5}$\\		
4451.0 &	2.5559 &	5.9826$\times 10^{-5}$\\		
\hline
\end{tabular}
\begin{tabular}{p{10cm}}
$^\ast$Table continued as electronic-only file.
\end{tabular}
\end{table}

\end{document}